\documentclass{JINST}
\usepackage{amsmath,amssymb,mathrsfs,graphicx}

\newtheorem{mytheo}{Theorem}[section]
\title{Analytical modeling of thin film neutron converters and its application to thermal neutron gas detectors}

\author{Francesco Piscitelli$^{a,b}$\thanks{Corresponding
author. }  and Patrick Van Esch$^a$\\
\llap{$^a$} Institut Laue-Langevin (ILL),\\ 6, Jules Horowitz, 38042
Grenoble, France. \\
\llap{$^b$} University of Perugia, \\
Piazza Universit\`a 1, 06123 Perugia, Italy.\\
E-mail: \email{piscitelli@ill.fr}}

\abstract{A simple model is explored mainly analytically to
calculate and understand the PHS of single and multi-layer thermal
neutron detectors and to help optimize the design in different
circumstances. Several theorems are deduced that can help guide the
design.}


\begin{document}

\section{Introduction}
\begin{figure}[ht!] \centering
\includegraphics[width=9cm,angle=0,keepaspectratio]{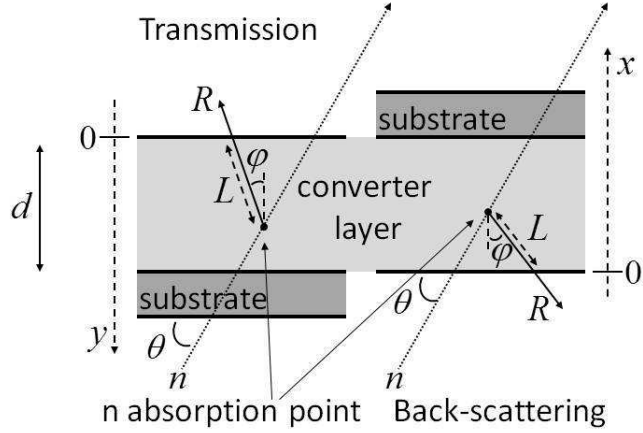}
\caption{\footnotesize Variables definition for a
\emph{back-scattering} and \emph{transmission} layer calculations.
\label{coorsys}}
\end{figure}

Using powerful simulation software has the advantage of including
many effects and potentially results in high accuracy. On the other
hand it does not always give the insight an equation can deliver.
\\ This paper originates from the necessity to understand the Pulse
Height Spectra (PHS) given by solid neutron converters employed in
thermal neutron detectors as in \cite{jonisorma},
\cite{kleinjalousie}, \cite{buff3}, \cite{lacy1}, and from the
investigation over such a detectors' efficiency optimization. Many
efforts have been recently made in order to address the $^3He$
shortage problem. The development of new technologies in neutron
detection is important for both national security \cite{kouzes2} and
for scientific research \cite{gebauer1}. Examples of application can
be found in \cite{athanasiades1}, \cite{gregor2} and
\cite{tsorbatzoglou1}.
\\ When a neutron is converted in a gaseous medium, such as a $^3He$
detector, the neutron capture reaction fragments ionize the gas
directly and the only energy loss is due to the wall effect. As a
result, such detectors show a very good gamma-rays to neutron
discrimination because gamma-rays release only a small part of their
energy in the gas volume and consequently neutron events and
gamma-rays events are easily distinguishable on the PHS. \\ On the
other hand, when dealing with hybrid detectors as in
\cite{jonisorma}, where the neutron converter is solid and the
detection region is gaseous, the gamma-ray to neutron discrimination
can be an issue \cite{lacy2}, \cite{athanasiades1}. Indeed once a
neutron is absorbed in the solid converter, it gives rise to charged
fragments which have to travel across part of the converter layer
itself before reaching the gas volume to originate a detectable
signal. As a result, those fragments can release only a part of
their energy in the gas volume. The neutron PHS can thus have
important low energy contributions, therefore gamma-ray and neutron
events are not well separated just in energy.
\\ In this paper we want to give a comprehension of the important aspects
of the PHS by adopting a simple theoretical model for solid neutron
converters. We will show good agreement of the model with the
measurements obtained on a $^{10}B$-based detector.
\\ In the same way the analytical model can help us optimize the
efficiency for single and multi-layer detectors in different
circumstances of incidence angle and neutron wavelength
distribution.
\\ The model we use is the same as implicitly used in many papers
such as \cite{gregor} or \cite{salvat}. It makes the following
simplifying assumptions:
\begin{itemize}
    \item the tracks of the emitted particles are straight lines
    emitted back-to-back and distributed isotropically;
    \item the energy loss is deterministic and given by the Bragg
    curves without fluctuations;
    \item the energy deposited is proportional to the charge
    collected without fluctuations.
\end{itemize}
Referring to Figure \ref{coorsys}, we talk about a
\emph{back-scattering} layer when neutrons are incident from the
gas-converter interface and the escaping particles are emitted
backwards into the gas volume; we call it a \emph{transmission}
layer when neutrons are incident from the substrate-converter
interface and the escaping fragments are emitted in the forward
direction in the sensitive volume. We consider a neutron to be
converted at certain depth ($x$ for back-scattering or $d-y$ for
transmission) in the converter layer and its conversion yields two
charged particles emitted back-to-back.

\section{Double layer}\label{Sect2laysub}
\begin{figure}[ht!]
\centering
\includegraphics[width=10cm,angle=0,keepaspectratio]{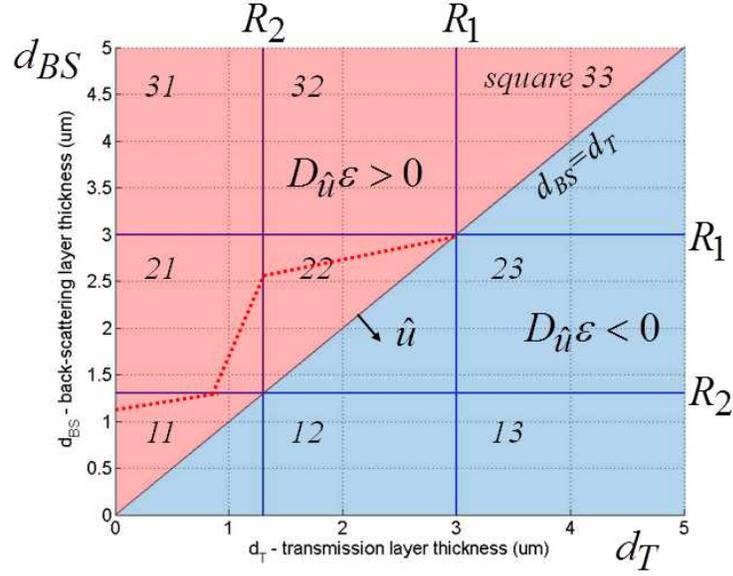}
\caption{\footnotesize Domain for the blade efficiency function. The
domain is divided into 9 partitions according to the neutron capture
fragment ranges. In red and blue are the domain partitions where the
efficiency directional derivative along the unity vector $\hat{u}$
is respectively always positive or always negative. The red dotted
line represents the case when the substrate effect is not negligible
and the maximum efficiency can not be attained on the domain
bisector.} \label{fig2laydom}
\end{figure}

We put a double coated blade in a gas detection volume. A
\emph{blade} consists of a substrate holding two converter layers,
one in back-scattering mode and one in transmission mode.
\\ Starting from the analytical formulae derived in \cite{gregor}
we are going to derive properties that can help to optimize the
efficiency in the case of a monochromatic neutron beam and in the
case of a distribution of neutron wavelengths. \\ By denoting with
$d_{BS}$ the thickness of the coating for the back-scattering layer
and with $d_{T}$ the transmission layer thickness, the efficiency of
the whole blade is:
\begin{equation}\label{eqac1}
\varepsilon(d_{BS},d_{T})=\varepsilon_{BS}\left(d_{BS}\right)+e^{-\Sigma
\cdot d_{BS}}\cdot \varepsilon_{T}\left(d_{T}\right)
\end{equation}
\\ where $\varepsilon_{BS}\left(d_{BS}\right)$ and
$\varepsilon_{T}\left(d_{T}\right)$ are the efficiency for a single
coating calculated as shown in appendix \ref{app1} from
\cite{gregor}, and $\Sigma$ is defined in Equation \ref{eqac2abc}.
The relation $\nabla \varepsilon(d_{BS},d_{T})=0$ determines the two
optimal layer thicknesses.
\\ In order to keep calculations simple, we consider only two
neutron capture fragments yielded by the reaction. This
approximation will not affect the meaning of the conclusion. In the
case of $^6Li$ Equation \ref{eqac1} is exact, for $^{10}B$ the
expression \ref{eqac1} should ideally be replaced by
$\varepsilon(d_{BS},d_{T})=0.94 \cdot\left(
\varepsilon^{0.94}_{BS}\left(d_{BS}\right)+e^{-\Sigma \cdot
d_{BS}}\cdot \varepsilon^{0.94}_{T}\left(d_{T}\right)\right)+0.06
\cdot\left( \varepsilon^{0.06}_{BS}\left(d_{BS}\right)+e^{-\Sigma
\cdot d_{BS}}\cdot \varepsilon^{0.06}_{T}\left(d_{T}\right)\right)$;
where $\varepsilon^{0.94}$ means the efficiency calculated for the
$94\%$ branching ratio reaction with the right effective particle
ranges. We will limit us to the $94\%$ contribution as if it were
$100\%$. $R_1$ and $R_2$, with ($R_2<R_1$), are the two ranges of
the two neutron capture fragments. In case of $^{10}B_4C$ the two
$94\%$ branching ratio reaction particle ranges are $R_1=3\,\mu m$
($\alpha$-particle) and $R_2=1.3\,\mu m$ ($^7Li$), when a $100\,KeV$
energy threshold is applied (as defined the minimum detectable
energy in \cite{gregor}).
\\ As $\varepsilon_{BS}\left(d\right)$ and
$\varepsilon_{T}\left(d\right)$ have different analytical
expressions according to whether $d \leq R_2<R_1$, $R_2<d \leq R_1$
or $R_2<R_1<d$ we need to consider 9 regions to calculate $\nabla
\varepsilon(d_{BS},d_{T})$ as shown in Figure \ref{fig2laydom}. If
we were to include the four different reaction fragments we would
have to consider $25$ domain partitions.
We will see later that the important regions, concerning the
optimization process, are the regions \emph{square 11} (where
$d_{BS} \leq R_2<R_1$ and $d_{T} \leq R_2<R_1$) and \emph{square 22}
(where $R_2<d_{T} \leq R_1$, $R_2<d_{BS} \leq R_1$).
\\ In order to consider a non-orthogonal incidence of
neutrons on the layers, it is sufficient to replace $\Sigma$ with
$\frac{\Sigma}{\sin(\theta)}$, where $\theta$ is the angle between
the neutron beam and the layer surface (see Figure \ref{coorsys}).
This is valid for both the single blade case and for a multi-layer
detector. The demonstration can be found in \cite{gregor}.

\subsection{Monochromatic double layer optimization}
In the domain region called \emph{square 11} the efficiency turns
out to be:
\begin{equation}\label{eqac2}
\varepsilon_{11}\left(d_{BS},d_{T}\right) = A \cdot
\left(1-e^{-\Sigma d_{BS}} \right)+d_{BS} \cdot C \cdot e^{-\Sigma
d_{BS}} + e^{-\Sigma d_{BS}} \left(B \cdot \left(1-e^{-\Sigma d_{T}}
\right)-C \cdot d_{T} \right)
\end{equation}
\\ Where we have called:
\begin{equation}\label{eqac2abc}
A=\left(1-\frac{1}{2\Sigma R_1}-\frac{1}{2\Sigma R_2} \right), \quad
B=\left(1+\frac{1}{2\Sigma R_1}+\frac{1}{2\Sigma R_2} \right), \quad
C=\left(\frac{1}{2 R_1}+\frac{1}{2 R_2} \right), \quad \Sigma= n
\cdot \sigma(\lambda)
\end{equation}
\\ where $n$ is the number density of the converter layer and
$\sigma(\lambda)$ its neutron absorption cross-section.
\\ By calculating $\nabla \varepsilon_{11}(d_{BS},d_{T})=0$
we obtain the result that $d_{BS} = d_{T}$ and
\begin{equation}\label{eqac3}
d_{BS} = d_{T} =-\frac{1}{\Sigma}\cdot \ln \left(\frac{C}{\Sigma
B}\right)
\end{equation}
\\ We repeat the procedure for the \emph{square 22} in which the efficiency is:
\begin{equation}\label{eqac4}
\begin{array}{ll}
\varepsilon_{22}\left(d_{BS},d_{T}\right)&= A +\frac{e^{-\Sigma
R_2}}{2 \Sigma R_2}-\frac{e^{-\Sigma d_{BS}}}{2}\cdot \left(1-
\frac{1}{\Sigma R_1}-\frac{d_{BS}}{R_1}\right)+
\\&
+e^{-\Sigma d_{BS}}\cdot \left(- B \cdot e^{-\Sigma
d_{T}}+\frac{1}{2} \left(1+\frac{1}{\Sigma R_1} -
\frac{d_{T}}{R_1}\right)+\frac{e^{-\Sigma \left(d_{T}-R_2
\right)}}{2\Sigma R_2} \right)
\end{array}
\end{equation}
\\ We obtain again $d_{BS} = d_{T}$ and
\begin{equation}\label{eqac5}
d_{BS} = d_{T} =-\frac{1}{\Sigma}\cdot \ln \left(\frac{R_2}{R_1}
\left(\frac{1}{2 R_2 \Sigma B - e^{+\Sigma R_2}}\right)\right)
\end{equation}
\\ Naturally each result of Equations \ref{eqac3} and \ref{eqac5}
is useful only if it gives a value that falls inside the region it
has been calculated for.
\\ The points defined by Equations \ref{eqac3} and \ref{eqac5} define
a \emph{maximum} of the efficiency function in the regions
\emph{square 11} or \emph{square 22} because the Hessian matrix in
those points has a positive determinant and $\frac{\partial^2
\varepsilon}{\partial d_{BS}^2}$ is negative.
\begin{figure}[!ht]
\centering
\includegraphics[width=7.5cm,angle=0,keepaspectratio]{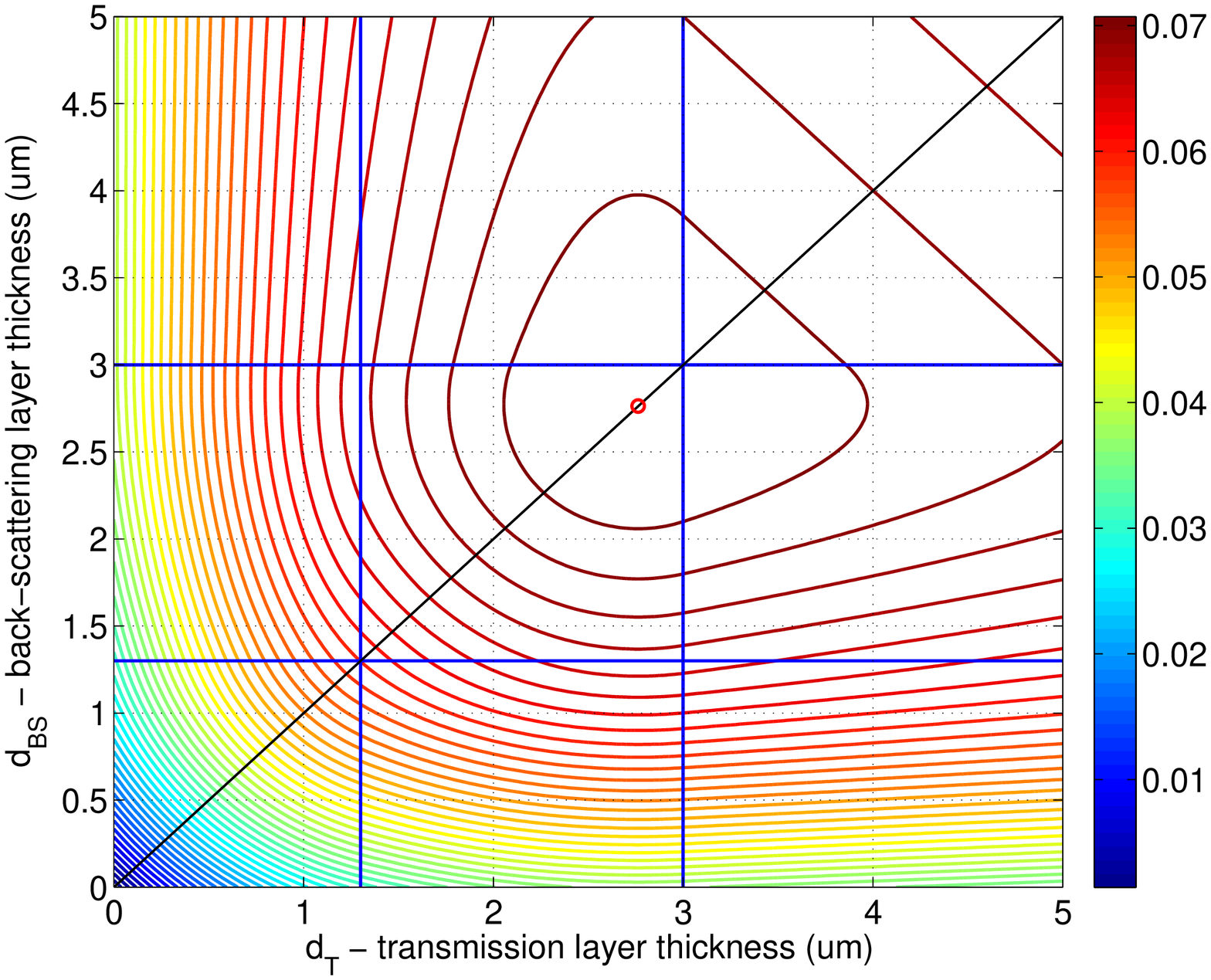}
\includegraphics[width=7.5cm,angle=0,keepaspectratio]{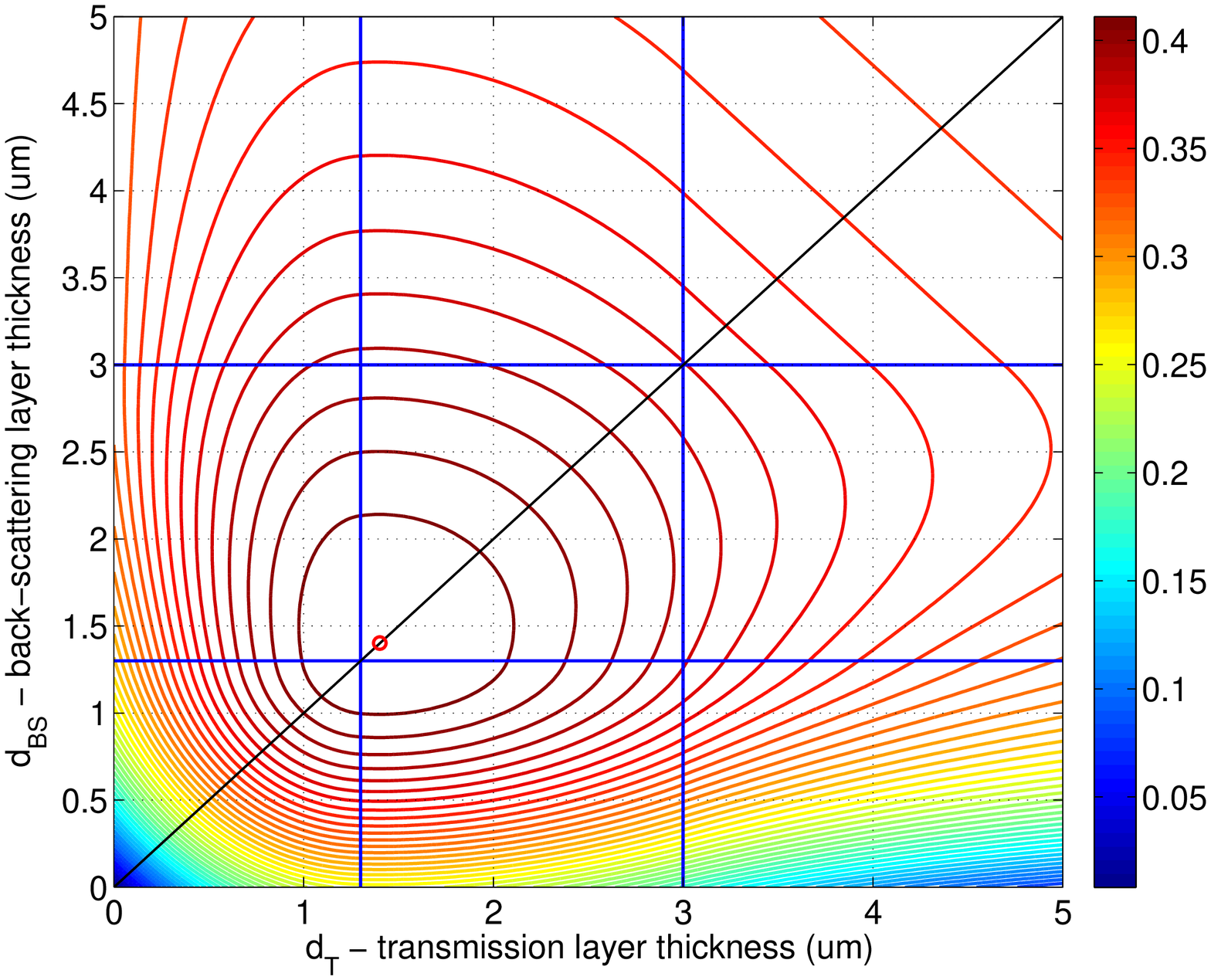}
 \caption{\footnotesize Efficiency plot for a double coated
substrate with $^{10}B_4C$ at $\theta=90{^\circ}$ incidence at
1.8\AA \, (left) and 20\AA \, (right).} \label{fig2lay1}
\end{figure}
\begin{figure}[!ht]
\centering
\includegraphics[width=7.5cm,angle=0,keepaspectratio]{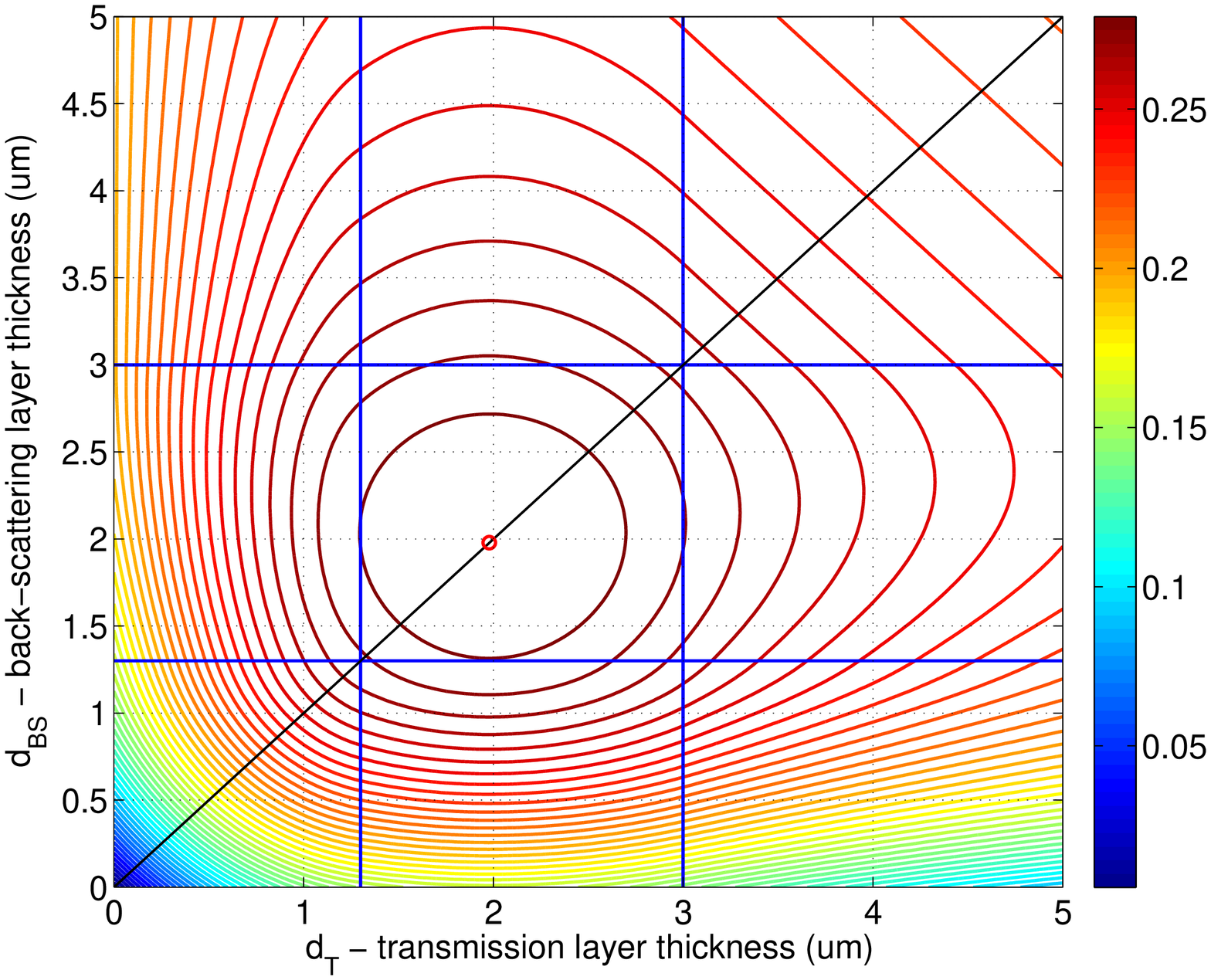}
\includegraphics[width=7.5cm,angle=0,keepaspectratio]{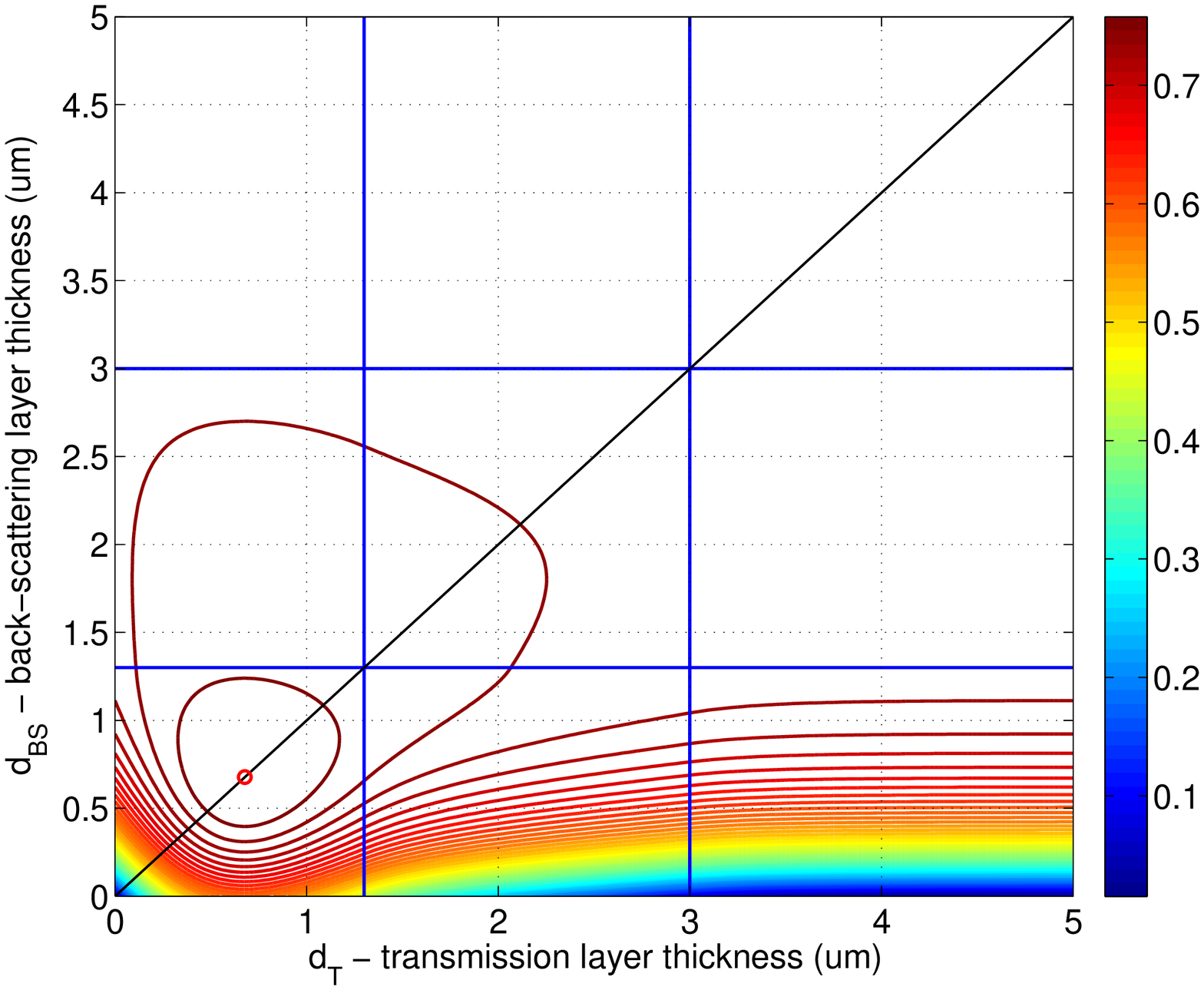}
 \caption{\footnotesize Efficiency plot for a double coated
substrate with $^{10}B_4C$ at $\theta=10{^\circ}$ incidence at
1.8\AA \, (left) and 20\AA \, (right).} \label{fig2lay2}
\end{figure}
It is easy to demonstrate there are no extreme points outside the
domain regions where either $d_{BS} > R_1$ or $d_{T} > R_1$, i.e.
\emph{square $jk$} with $j=3$ or $k=3$ or both. This outcome is also
intuitive. In back-scattering mode when the converter thickness
becomes thicker than the longest particle range ($R_1$) there is no
gain in efficiency by adding more converter material. In the
transmission case, increasing the thickness above $R_1$ will add
material that can only absorb neutrons without any particle
escaping.
\\ For the cases of \emph{squares 12} and
\emph{21}, we obtain that $\nabla \varepsilon(d_{BS},d_{T})=0$ has
no solution; thus the efficiency maximum can never fall in these
domain regions.
\\ In Figures \ref{fig2lay1} and \ref{fig2lay2} the efficiency for
four different cases is plotted. The red circle identifies the point
of maximum efficiency calculated by using Equations \ref{eqac3} or
\ref{eqac5}; it stands out immediately that even though the
efficiency function is not symmetric relatively to the domain
bisector (drawn in black) the maximum nevertheless always lies on
it.
\\ \emph{This is a very important result because the sputtering deposition
method \cite{carina} coats both sides of the substrate with the same
thickness of converter material and it is also suited to make
optimized blades}.

\subsection{Effect of the substrate}
If we consider the neutron loss due to the substrate, the Equation
\ref{eqac1} has to be modified as follows:
\begin{equation}\label{eqac1bis}
\varepsilon(d_{BS},d_{T})=\varepsilon_{BS}\left(d_{BS}\right)+\delta(\lambda)
\cdot e^{-\Sigma \cdot d_{BS}}\cdot
\varepsilon_{T}\left(d_{T}\right)
\end{equation}
\\ Where $\delta(\lambda)=e^{-\Sigma_{sub}(\lambda) \cdot d_{sub}}$
and $\Sigma_{sub}$ and $d_{sub}$ are the macroscopic cross-section
and the thickness of the substrate. \\ If we optimize the layer
thicknesses, we find the same result of Equations \ref{eqac3} and
\ref{eqac5} for the transmission layer thickness $d_T$ but, on the
other hand, the back-scattering layer thickness does not equal the
transmission layer thickness anymore. It becomes:
\begin{equation}\label{eqau1122}
d_{BS} = \begin{cases} \frac{1}{C} \cdot \left(1-\delta\right)+\delta \cdot d_{T} &\mbox{for \emph{square 11}}\\
R_1 \cdot\left(1-2\delta\right)+\delta \cdot
\left(1+\frac{R_1}{R_2}\right) \cdot d_{T} &\mbox{for \emph{square
21}}\\
R_1 \cdot\left(1-\delta\right)+\delta \cdot d_{T} &\mbox{for
\emph{square 22}}
\end{cases}
\end{equation}
\\ The maximum efficiency can now also lie in \emph{square 21}
but not in \emph{square 12}, because $\delta \in [0,1]$. The dotted
line in \emph{squares 11, 21} and \emph{22}, in Figure
\ref{fig2laydom}, are Equations \ref{eqau1122}. The slope of the
line in \emph{squares 11} and \emph{22} is equal to
$\delta(\lambda)<1$. The maximum efficiency, when $\delta(\lambda)$
is not negligible, now lies on the dotted line just above the
optimum without substrate effect.
\\ From Equation \ref{eqau1122} we can observe that when $\delta$ is
close to zero, i.e. the substrate is very opaque to neutrons, the
thickness of the back-scattering layer tends to the value $R_1$. On
the other hand, when $\delta$ is close to one, $d_{BS}$ tends to
$d_{T}$. The factor $\delta(\lambda)$ is usually very close to one
for many materials that serve as substrate. We define the relative
variation between $d_{BS}$ and $d_{T}$ as $\Delta_{d} =\left|
\frac{d_{T}-d_{BS}}{d_{T}}\right|$; in Table \ref{tabdelta99} they
are listed for a $0.5\,mm$ and $3\, mm$ Aluminium substrate of
density $\rho=2.7\, g/cm^3$ at $1.8$\AA. In case the substrate is
inclined under an angle of $10^{\circ}$ a substrate of $0.5\,mm$
looks like a substrate of about $3\, mm$. We consider a neutron to
be lost when it is either scattered or absorbed, therefore, the
cross-section used \cite{sears} is:
$\sigma_{Al}=\sigma_{Al}^{abs}(\lambda)+\sigma_{Al}^{scatt}= 0.2\,b
$(at $ 1.8$\AA )$+1.5\,b=1.7\, b$.

\begin{table}[!ht]
\caption{\footnotesize Neutron loss factor $\delta$ for an Aluminium
substrate and relative difference between the two coating
thicknesses held by the substrate for $1.8$\AA.} \centering
\begin{tabular}{|c|c|c|}
\hline \hline
$d_{sub}(mm)$ &  $\delta$ (1.8\AA) &$\Delta_{d}$ (1.8\AA) \\
\hline
0.5 &  0.995 & 0.0004 \\
3   &  0.970 & 0.0026  \\
\hline \hline
\end{tabular}
\label{tabdelta99}
\end{table}

\subsection{Double layer for a distribution of neutron
wavelengths}\label{dlfadnw1} The result of having the same optimal
coating thickness for each side of a blade is demonstrated for
monochromatic neutrons and we want to prove it now for a more
general case when the neutron beam is a distribution of wavelengths
and when the substrate effect can be neglected
($\delta(\lambda)\approx1$).
\\ We will prove a property that will turn out to be useful. We will
show that the directional derivative of $\varepsilon(d_{BS},d_{T})$
along the unit vector $\hat{u}=\frac{1}{\sqrt{2}}\left(1,-1\right)$
is positive until the bisector $d_{BS} = d_{T}$ and it changes sign
only there. This vector identifies the orthogonal direction to the
bisector (see Figure \ref{fig2laydom}).
\\ In the \emph{square 11}:
\begin{equation}\label{eqac6}
D_{\hat{u}}\varepsilon = \hat{u} \times
\nabla\varepsilon(d_{BS},d_{T}) = \frac{C \Sigma}{\sqrt{2}}\cdot
e^{-\Sigma d_{BS}} \left( d_{BS}-d_{T} \right)
\end{equation}
\\ In the \emph{square 22}:
\begin{equation}\label{eqac7}
D_{\hat{u}}\varepsilon= \frac{\Sigma}{2\sqrt{2}R_1}\cdot e^{-\Sigma
d_{BS}} \left( d_{BS}-d_{T} \right)
\end{equation}
\\ Which are both positive above the bisector and negative below.
In the other domain regions the demonstration is equivalent. E.g. in
the \emph{square 12}:
\begin{equation}\label{eqac8}
D_{\hat{u}}\varepsilon= -\frac{\Sigma}{2\sqrt{2}}\cdot e^{-\Sigma
d_{BS}} \left( 1 - \frac{d_{BS}}{R_2}- \frac{d_{BS}}{R_1} +
\frac{d_{T}}{R_1} \right)
\end{equation}
\\ Which is strictly negative in the \emph{square 12} where $d_{BS} \leq
R_2<R_1$ and $ R_2<d_{T} \leq R_1$ except in the corner, on the
bisector, where $d_{BS}=d_T=R_2$.
\\ The following theorem is therefore proved.
\begin{mytheo}\label{theo1}
\emph{The efficiency function defined by the Equation \ref{eqac1} is
strictly monotone in the two half-domains identified by the bisector
$d_{BS} = d_{T}$ (see Figure \ref{fig2laydom}).}
\end{mytheo}
In a general instrument design one can be interested in having a
detector response for a whole range of $\lambda$. E.g. an elastic
instrument can work a certain time at one wavelength and another
time at another wavelength. In a Time-Of-Flight instrument one can
be interested in having a sensitivity to neutrons of a certain
energy range including or excluding the elastic peak. One can define
a normalized weight function $w\left(\lambda \right)$
($\int_{0}^{+\infty}w\left(\lambda \right)\, d\lambda=1$) that
signifies how much that neutron wavelength is important compared to
others. I.e. the price we want to spend in a neutron scattering
instrument to be able to detect a neutron energy with respect to an
other one. We can also consider an incident beam of neutrons, whose
wavelength distribution is $w\left(\lambda \right)$, and we want to
maximize the efficiency given this distribution.
\\ The efficiency for a blade exposed to a neutron flux which shows
this distribution is:
\begin{equation}\label{eqac9}
\varepsilon_w (d_{BS},d_{T}) = \int_{0}^{+\infty}w\left(\lambda
\right) \varepsilon (d_{BS},d_{T},\lambda) \, d\lambda
\end{equation}
\\ where $\varepsilon (d_{BS},d_{T},\lambda)$ is the efficiency in
Equation \ref{eqac1}.
\\ In order to optimize this efficiency its gradient
relative to $d_{BS}$ and $d_{T}$ has to be calculated:
\begin{equation}\label{eqac10}
\nabla\varepsilon_w (d_{BS},d_{T}) =\int_{0}^{+\infty}w\left(\lambda
\right) \nabla\varepsilon (d_{BS},d_{T},\lambda) \, d\lambda
\end{equation}
Both gradient components have to cancel out ($\frac{\partial
\varepsilon_w}{\partial d_{BS}} =\frac{\partial
\varepsilon_w}{\partial d_{T}}=0$), this leads to
$D_{\hat{u}}\varepsilon_w=0$. E.g. in \emph{square 11}:
$\int_{0}^{+\infty}w\left(\lambda \right)\left(e^{-\Sigma d_{BS}}
\left(d_T-d_{BS}\right) C \Sigma \right)\, d\lambda=0$. As a result,
in order for the efficiency to attain a maximum, it is necessary
(but not sufficient) that its directional derivative along the unity
vector $\hat{u}=\frac{1}{\sqrt{2}}\left(1,-1\right)$:
\begin{equation}\label{eqac12}
D_{\hat{u}}\varepsilon_w=\int_{0}^{+\infty}w\left(\lambda
\right)D_{\hat{u}}\varepsilon(d_{BS},d_{T},\lambda)\, d\lambda
\end{equation}
equals zero.
\\ For a general family of functions $f(d_{BS},d_{T},\lambda)$
for which the maximum always lies on the domain bisector it is not
true that the function defined by their positively weighted linear
combination must have the maximum on $d_{BS}=d_{T}$, because in
general $\nabla f (d_{BS},d_{T},\lambda)$ can be positive, null or
negative, thus there are many ways to accomplish $\nabla f_w
(d_{BS},d_{T})=0$. However, thanks to Theorem \ref{theo1},
$D_{\hat{u}}\varepsilon=0$ is satisfied only on the bisector. Below
the bisector, this is always negative, as it is a positively
weighted integral of negative values; similarly, above the bisector,
this is always positive. Hence, the maximum can only be attained on
the bisector.
\\ The gradient can hence be replaced by
$\frac{\partial }{\partial d_T}$ and the function maximum has to be
searched on the bisector, therefore:
\begin{equation}\label{eqac11} \nabla\varepsilon_w
(d_{T}) =\int_{0}^{+\infty}w\left(\lambda \right)
\frac{\partial}{\partial d_T}\varepsilon
(d_{BS}=d_{T},d_{T},\lambda) \, d\lambda
\end{equation}
\\ \emph{In the end, the same layer thickness for both sides
of a blade has to be chosen in order to maximize its efficiency,
even if it is exposed to neutrons belonging to a general wavelength
distribution $w\left(\lambda \right)$}.
\\ The integration over $\lambda$ can be alternatively executed in
the variable $\Sigma$; indeed $\Sigma$ is just a linear function in
$\lambda$ because $\sigma_{abs}$ is proportional to $\lambda$ in the
thermal neutron region. Moreover, as indicated previously, $\Sigma$
is also a function of $\theta$ and this is the only appearance of
$\theta$ in the efficiency function. Hence, we can just as well
consider a weighting in $\lambda$ and $\theta$ which results in just
a weighting function over $\Sigma$. In other words, all the results
that have been derived for a wavelength distribution also hold for
an angular distribution or both.
\subsubsection{Flat neutron wavelength distribution example}
As a simple example we take a flat distribution between two
wavelengths $\lambda_1$ and $\lambda_2$ defined as follows:
\begin{equation}\label{eqac13}
w\left(\lambda\right)=\frac{1}{\lambda_2-\lambda_1}
\end{equation}
\\ In the \emph{square 11} we obtain:
\begin{equation}\label{eqac14}
\frac{\partial}{\partial d_T}\varepsilon
(d_{BS}=d_{T},d_{T},\lambda)=2e^{-\Sigma d_T}\left( B \Sigma
e^{-\Sigma d_T}-C\right)
\end{equation}
\\ We call $\Sigma_1=\Sigma(\lambda_1)$ and
$\Sigma_2=\Sigma(\lambda_2)$. We recall that A and B are function of
$\Sigma(\lambda)$.
\begin{equation}\label{eqac15}
\nabla\varepsilon_w (d_{T})
=\frac{1}{(\Sigma_2-\Sigma_1)}\left[\frac{e^{-2\Sigma
d_T}}{d_T}\cdot\left( 2Ce^{+\Sigma
d_T}-C-\Sigma-\frac{1}{2d_T}\right) \right]_{\Sigma_1}^{\Sigma_2}=0
\end{equation}
\\ In the same way the solution in the \emph{square
22} can be determined.
\begin{equation}\label{eqac17}
\frac{\partial \varepsilon}{\partial d_T} =e^{-\Sigma d_T}\left(
e^{-\Sigma d_T}\left(2 B \Sigma-\frac{e^{+\Sigma R_2}}{R_2}
\right)-\frac{1}{R_1}\right)
\end{equation}
\\ By integrating we finally obtain:
\begin{equation}\label{eqac16}
\nabla\varepsilon_w (d_{T})
=\frac{1}{(\Sigma_2-\Sigma_1)}\left[\frac{e^{-2 \Sigma
d_T}}{d_T}\left( \frac{e^{+\Sigma d_T}}{R_1}-C-\Sigma-\frac{1}{2
d_T}- \frac{d_T \, e^{+\Sigma R_2}}{R_2\left(R_2-2d_T
\right)}\right) \right]_{\Sigma_1}^{\Sigma_2}=0
\end{equation}
The solution of Equations \ref{eqac15} and \ref{eqac16} gives the
optimum value for the thickness of the two converter layers in the
region of the domain called \emph{square 11} and \emph{square 22}
respectively for a uniform neutron wavelength distribution between
$\lambda_1$ and $\lambda_2$. E.g. for a uniform neutron wavelength
distribution between $1$\AA \, and $20$\AA \, the optimal thickness
of coatings on both sides of the blade is $1\, \mu m$.

\section{The multi-layer detector design}
\begin{figure}[!ht]
\centering
\includegraphics[width=8cm]{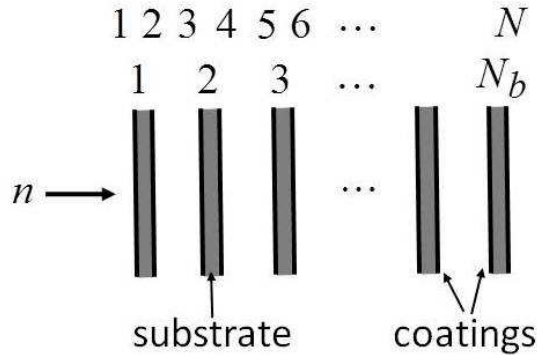}
\caption{\footnotesize A multi-layer detector schematic. $N_b$
blades (holding $N=2\cdot N_b$ converter layers) are placed in
cascade, alternating with detection regions.} \label{multigridschet}
\end{figure}
In a detector like that presented in \cite{jonisorma}, \cite{buff2},
or in \cite{wang1}, all the substrates have the same coating
thickness. One can ask if it is possible to optimize the coating
thicknesses for each layer in order to gain in efficiency. This is
also applicable to neutron detectors which use solid converters
coupled with GEMs \cite{kleincascade}. In a multi-layer detector
(see Figure \ref{multigridschet}), composed by $N$ layers or
$N_b=\frac{N}{2}$ blades, the efficiency can be written as follows:
\begin{equation}\label{eqad1}
\varepsilon_{tot}(N_b) = \varepsilon_1(d_{BS1},d_{T1})+
\sum_{k=2}^{N_b}\varepsilon_1(d_{BSk},d_{Tk})\cdot e^{- \left(
\sum_{j=1}^{\left(k-1\right)} \left(d_{BSj}+d_{Tj}\right)
\right)\cdot \Sigma }
\end{equation}
\\ Where $\varepsilon_1(d_{BSk},d_{Tk})=\varepsilon_{1k}$ represents the efficiency
for a single blade already defined by the Equation \ref{eqac1};
$d_{BSk}$ and $d_{Tk}$ are the coating thicknesses of the $k-th$
blade. \\ If the detector is assembled with blades of the same
thickness, i.e. $d_{BSk}=d_{Tk}=d, \,\, \forall k=1,2,\dots,N_b$,
Equation \ref{eqad1} can be simplified as follows:
\begin{equation}\label{eqad1sthick}
\varepsilon_{tot}(N_b) =\varepsilon_1(d) \cdot \sum_{k=1}^{N_b}
e^{-2 \left( k-1 \right) d \cdot \Sigma }=\varepsilon_1(d) \cdot
\frac{1-e^{-2 d \Sigma N_b}}{1-e^{-2 d \Sigma }}
\end{equation}
Therefore, $\frac{d \varepsilon_{tot}}{dd}=0$ has to be calculated
in order to optimize such a detector containing blades of same
coating thicknesses.
\\In the case of a distribution of wavelengths,
defined by $w\left(\lambda \right)$, Equation \ref{eqad1} has to be
integrated over $\lambda$ as already shown in Section
\ref{dlfadnw1}.
\begin{equation}\label{eqad14}
\varepsilon_{tot}^w (N_b,\bar{d}_{BS},\bar{d}_{T}) =
\int_{0}^{+\infty}w\left(\lambda \right)
\varepsilon_{tot}(N_b,\lambda) \, d\lambda
\end{equation}
The efficiency will be function of $N=2\cdot N_b$ variables; which
can be denoted using the compact vectorial notation by the two
vectors $\bar{d}_{BS}$ and $\bar{d}_{T}$ of $N_b$ components each.
\\ Both for monochromatic mode and for a distribution the detector
optimization implies the calculation of the $N$-dimensional gradient
$\nabla\varepsilon_{tot}$, because in the case of a distribution the
gradient can be carried inside the integral over $\lambda$. We will
use the shorthand
$\varepsilon_{1k}=\varepsilon_{1}\left(d_{BSk},d_{Tk}\right)$.
\\ The $k-th$ $N$-dimensional gradient component for
back-scattering mode results to be:
\begin{equation}\label{eqad16}
\frac{\partial \varepsilon_{tot}}{\partial d_{BSk}} =
\begin{cases}
\frac{\partial \varepsilon_{1k}}{\partial d_{BSk}}-\Sigma \cdot
\sum_{p=\left(k+1\right)}^{N_b} \varepsilon_{1p} \cdot e^{- \left(
\sum_{i=1}^{\left(p-1\right)} \left(d_{BSi}+d_{Ti}\right)
\right)\cdot \Sigma } &\mbox{if \,} k = 1 \\ \frac{\partial
\varepsilon_{1k}}{\partial d_{BSk}}\cdot e^{- \left(
\sum_{j=1}^{\left(N_b-1\right)} \left(d_{BSj}+d_{Tj}\right)
\right)\cdot \Sigma }-\Sigma \cdot \sum_{p=\left(k+1\right)}^{N_b}
\varepsilon_{1p} \cdot e^{- \left( \sum_{i=1}^{\left(p-1\right)}
\left(d_{BSi}+d_{Ti}\right) \right)\cdot \Sigma} & \mbox{if \, }
1<k<N_b
\\ \frac{\partial \varepsilon_{1k}}{\partial d_{BSk}}\cdot
e^{- \left( \sum_{j=1}^{\left(N_b-1\right)}
\left(d_{BSj}+d_{Tj}\right) \right)\cdot \Sigma } & \mbox{if \, } k
= N_b
\end{cases}
\end{equation}
\\ Equivalently we find the same formal expression for the $k-th$ component
of the gradient with respect to the transmission variable ($d_T$);
we can substitute $\partial d_{BSk}$ with $\partial d_{Tk}$ in
Equation \ref{eqad16} which are for the rest entirely the same. This
implies that if we put identical conditions on $\frac{\partial
\varepsilon_{tot}}{\partial d_{BSk}}$ and on $\frac{\partial
\varepsilon_{tot}}{\partial d_{Tk}}$ this will result in
$\frac{\partial \varepsilon_{1k}}{\partial d_{BS}}=\frac{\partial
\varepsilon_{1k}}{\partial d_{T}}$. Equivalently as already found in
Section \ref{dlfadnw1}, from Equation \ref{eqad16} and the one for
the transmission variable we finally obtain ($\forall k
=1,2,\dots,N_b$):
\begin{equation}\label{eqad17}
\begin{cases}
\frac{\partial \varepsilon_{1k}}{\partial d_{BSk}}=\frac{\partial
\varepsilon_{1k}}{\partial d_{Tk}}\Rightarrow
D_{\hat{u}}\varepsilon_{1k}=0
&\mbox{if \,} k < N_b \\
\frac{\partial \varepsilon_{1k}}{\partial d_{BSk}}=\frac{\partial
\varepsilon_{1k}}{\partial d_{Tk}}=0 &\mbox{if \, } k = N_b
\end{cases}
\end{equation}
\\ Both for the monochromatic case and in the case of a distribution of wavelengths,
the condition in Equation \ref{eqad17} has to be satisfied. Using
Theorem \ref{theo1}, the maximum efficiency can only be found,
again, on the bisector.
\\ \emph{In a multi-layer detector, which has to
be optimized for any distribution of neutron wavelengths or for a
single wavelength, all the blades have to hold two layers of the
same thickness. Naturally, thicknesses of different blades can be
distinct}.
\\ Thanks to this property, we can denote with $d_k$
the common thickness of the two layers held by the $k-th$ blade
($d_{BSk}=d_{Tk}=d_k$), furthermore, Equation \ref{eqad1} can be
simplified as follows:
\begin{equation}\label{eqad7}
\varepsilon_{tot}(N,\bar{d})=\varepsilon_1(d_{1})+
\sum_{k=2}^{N_b}\varepsilon_1(d_{k})\cdot e^{-2 \left(
\sum_{j=1}^{\left(k-1\right)} d_j \right)\cdot \Sigma }
\end{equation}
where $\bar{d}$ is the vector of components $d_k$ for $k=1,2,\dots
,N_b$.
\\ Optimizing a detector for a single neutron wavelength or for a distribution is
different; the equation $\nabla\varepsilon_{tot}=0$ in one case and
$\nabla\varepsilon_{tot}^w=0$ in the other represent a
$N_b$-dimensional system of equations in $N_b$ unknowns because of
the simplification of having the same back-scattering and
transmission layer thickness on the blades. By expanding the
Equation \ref{eqad7} we obtain:
\begin{equation}\label{eqad2}
\begin{aligned}
\varepsilon_{tot}(N_b)&= \varepsilon_1(d_{1})+e^{-2
d_{1}\Sigma}\cdot \varepsilon_1(d_2)+ e^{-2 d_{1}\Sigma}\cdot e^{-2
d_{2}\Sigma} \cdot\varepsilon_1(d_{3})+\dots
\\& \dots+e^{-2d_{1}\Sigma}\cdot \dots \cdot
e^{-2 d_{BS(N_b-1)}\Sigma}\cdot
\varepsilon_1(d_{N_b})=\\&=\varepsilon_1(d_{1})+e^{-2d_{1}\Sigma}\cdot
\left[\varepsilon_1(d_{2})+e^{-2d_{2}\Sigma}\cdot \left[\dots
\left[\varepsilon_1(d_{(N_b-1)})
+e^{-2d_{(N_b-1)}\Sigma}\cdot\varepsilon_1(d_{N_b})
\right]\dots\right]\right]
\end{aligned}
\end{equation}
We notice that the variable $d_{N_b}$ appears only once, this means
that, in the case of a single wavelength, its value can be
determined without taking the others into account. Continuing the
reasoning we see that the system of equations is upper triangular.
In the monochromatic case we can optimize the detector starting from
the last blade and going backward till the first. This is not true
for the distribution case in which the gradient of Equation
\ref{eqad2} is in addition integrated over $\lambda$, thus all the
blades have to be taken into account simultaneously in the
optimization process. In the monochromatic case, we can start by
fixing the last blade coating thickness because any change on the
previous will only affect the \emph{number} of neutrons that reach
the last blade, and we require the last blade to be as efficient as
possible for that kind of neutron. As the layer thickness optimum of
each blade does not depend on the previous ones, the system is
triangular. On the other hand, in the case of a wavelength
distribution, any change on the previous blades will change the
actual \emph{distribution} of wavelengths the last blade
experiences. Thus, the neutron distribution a blade has to be
optimized for depends on all the previous blade coatings. In this
case, the system is not triangular.

\subsection{Monochromatic multi-layer detector optimization}
In order to optimize the layers in multi-layer detectors for a given
neutron wavelength, we can maximize the last layer efficiency and,
then, go backward until the first layer. Formally, from Equation
\ref{eqad2}, we obtain an iterative structure:
\begin{equation}\label{eqad3}
f_{k} = \begin{cases} \varepsilon_1(d_{k})+e^{-2d_{k}\Sigma}\cdot
\alpha_{k+1} &\mbox{if \,} k < N_b \\ \varepsilon_1(d_{k}) &\mbox{if
\, } k = N_b
\end{cases}
\end{equation}
$\alpha_{k+1}$ is a fixed number, independent from $d_k$, and
represents the cumulative efficiency of the detector from the blade
$(k+1)-th$ to the end.
\begin{equation}\label{eqad8}
\frac{d f_{k}}{d d_k} = \begin{cases}\frac{d }{d d_{k}}
\varepsilon_1(d_k)-2\Sigma\,
e^{-2 d_k \Sigma}\cdot \alpha_{k+1} &\mbox{if \,} k < N_b \\
\frac{d}{d d_{k}}\varepsilon_1(d_k) &\mbox{if \, } k = N_b
\end{cases}
\end{equation}
\\ $\frac{d }{d d_{k}}\varepsilon_1(d_k)$ are the derivatives in Equations \ref{eqac14} and
\ref{eqac17} according to the domain partitions. In the domain
region called \emph{square 11} as defined in Section
\ref{Sect2laysub}, we obtain:
\begin{equation}\label{eqad9}
\frac{d f_{k}}{d d_k} = \begin{cases}2e^{-\Sigma d_k}\left(
\left(B-\alpha_{k+1}\right) \Sigma e^{-\Sigma
d_k}-C\right)&\mbox{if \,} k < N_b \\
2e^{-\Sigma d_k}\left( B \Sigma e^{-\Sigma d_k}-C\right) &\mbox{if
\, } k = N_b
\end{cases}
\end{equation}
And in the \emph{square 22}:
\begin{equation}\label{eqad10}
\frac{d f_{k}}{d d_k} = \begin{cases}e^{-\Sigma d_k}\left(
e^{-\Sigma d_k}\left(2 \left(B -
\alpha_{k+1}\right)\Sigma-\frac{e^{+\Sigma R_2}}{R_2}
\right)-\frac{1}{R_1}\right)  &\mbox{if \,} k < N_b \\
e^{-\Sigma d_k}\left( e^{-\Sigma d_k}\left(2 B
\Sigma-\frac{e^{+\Sigma R_2}}{R_2} \right)-\frac{1}{R_1}\right)
&\mbox{if \, } k = N_b
\end{cases}
\end{equation}
\\ Equations \ref{eqad9} and \ref{eqad10} have solutions similar to
\ref{eqac3} and \ref{eqac5}. In the \emph{square 11} the solution
is:
\begin{equation}\label{eqad11}
d_k^{opt} = \begin{cases} -\frac{1}{\Sigma} \cdot \ln
\left(\frac{C}{\left(B-\alpha_{k+1}\right)\Sigma}\right) &\mbox{if \,} k < N_b \\
-\frac{1}{\Sigma} \cdot \ln \left(\frac{C}{B\Sigma}\right) &\mbox{if
\, } k = N_b
\end{cases}
\end{equation}
In the \emph{square 22}:
\begin{equation}\label{eqad12}
d_k^{opt} = \begin{cases} -\frac{1}{\Sigma} \cdot \ln
\left(\frac{R_2}{R_1}\left(\frac{1}{2 R_2 \Sigma \left(B-\alpha_{k+1}\right)- e^{+\Sigma R_2}}\right)\right) &\mbox{if \,} k < N_b \\
-\frac{1}{\Sigma} \cdot \ln \left(\frac{R_2}{R_1} \left(\frac{1}{2
R_2 \Sigma B - e^{+\Sigma R_2}}\right)\right) &\mbox{if \, } k = N_b
\end{cases}
\end{equation}
\\ The optimization method is a recursive procedure that employs the
Equations \ref{eqad11} and \ref{eqad12}; we start from the last
blade, and we find its optimal thickness $d_{N_b}^{opt}$, afterwards
we calculate $\alpha_{N_b}$ as the last layer efficiency using the
optimal thickness found. Now we can calculate $d_{N_b-1}^{opt}$ from
Equations \ref{eqad11} or \ref{eqad12} and $\alpha_{N_b-1}$ and so
on until the first layer.
\begin{equation}\label{eqad13}
\alpha_{k+1} = \begin{cases}
\varepsilon_1(d_{k+1}^{opt})+\sum_{i=k+2}^{N_b}\varepsilon_1(d_{i}^{opt})
\cdot e^{-2 \left(\sum_{j=k+1}^{\left(i-1\right)} d_{j}^{opt}
\right)\cdot \Sigma } &\mbox{if \,} k+1 < N_b
 \\ \varepsilon_1(d_{k+1}^{opt}) &\mbox{if \, } k+1 = N_b
\end{cases}
\end{equation}
\subsubsection{Example of application}
\begin{figure}[!ht]
\centering
\includegraphics[width=7.5cm,angle=0,keepaspectratio]{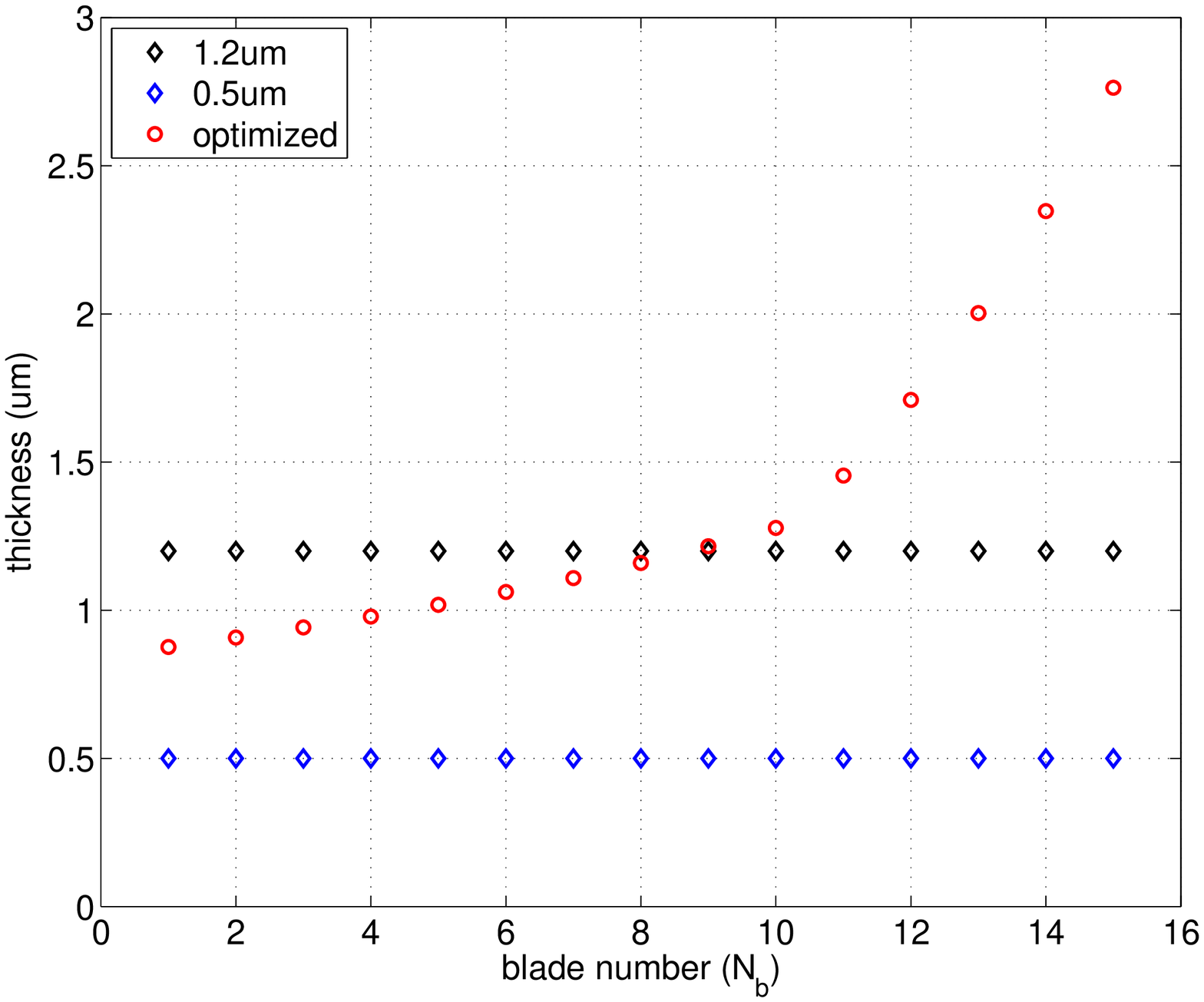}
\includegraphics[width=7.5cm,angle=0,keepaspectratio]{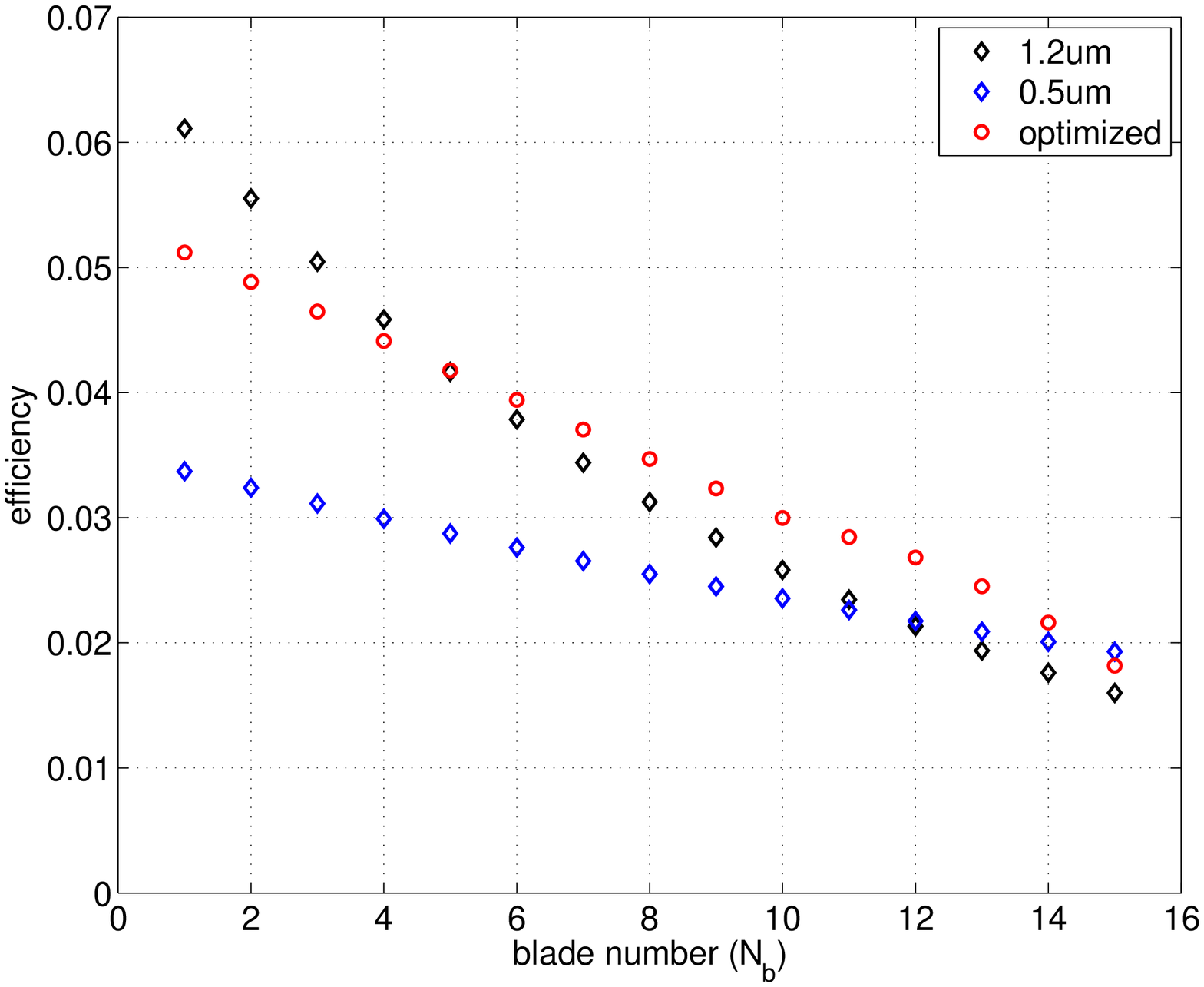}
 \caption{\footnotesize Thicknesses of the blade coatings (left) and their efficiency contribution (right),
 for a detector made up of $15$ blades
 of $1.2\, \mu m$, $0.5\, \mu m$ and for a detector optimized for $1.8$\AA.} \label{figMG30mon1p8}
\end{figure}

\begin{figure}[!ht]
\centering
\includegraphics[width=7.5cm,angle=0,keepaspectratio]{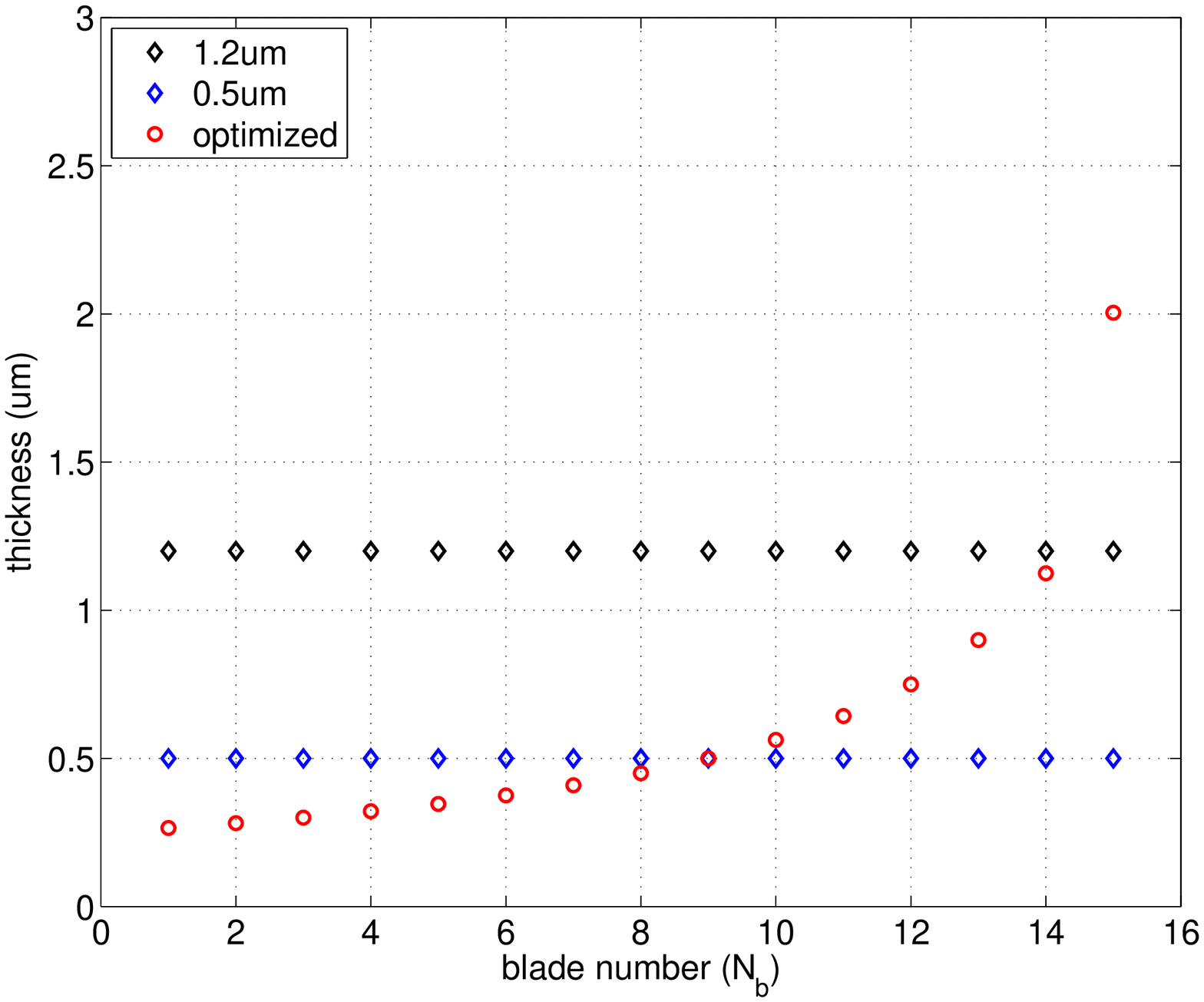}
\includegraphics[width=7.5cm,angle=0,keepaspectratio]{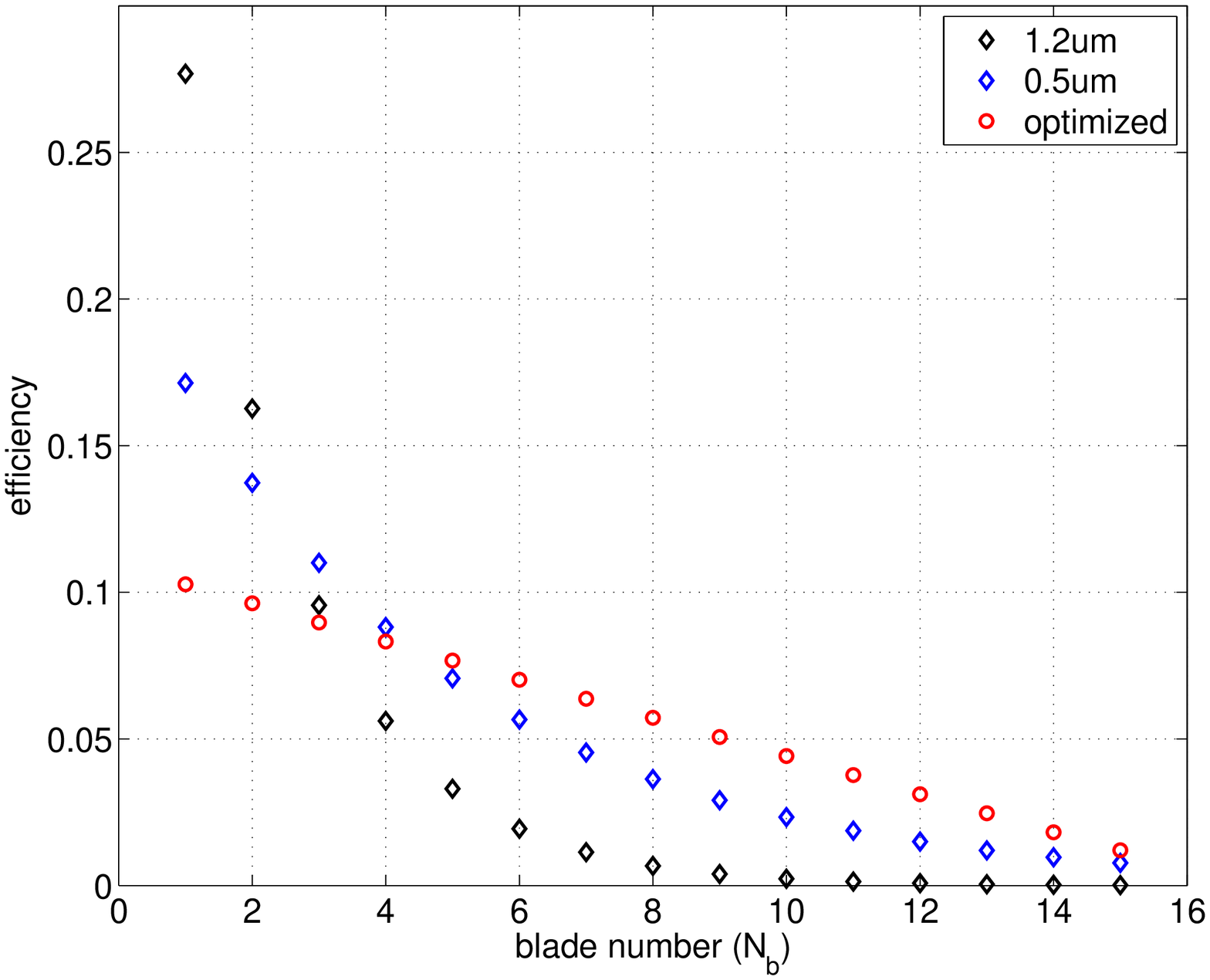}
 \caption{\footnotesize Thicknesses of the blades coatings (left) and their efficiency contribution (right),
 for a detector made up of $15$ identical coating thickness blades
 of $1.2\, \mu m$, $0.5\, \mu m$ and for a detector optimized for $10$\AA.} \label{figMG30mon20}
\end{figure}

\begin{figure}[!ht]
\centering
\includegraphics[width=7.5cm]{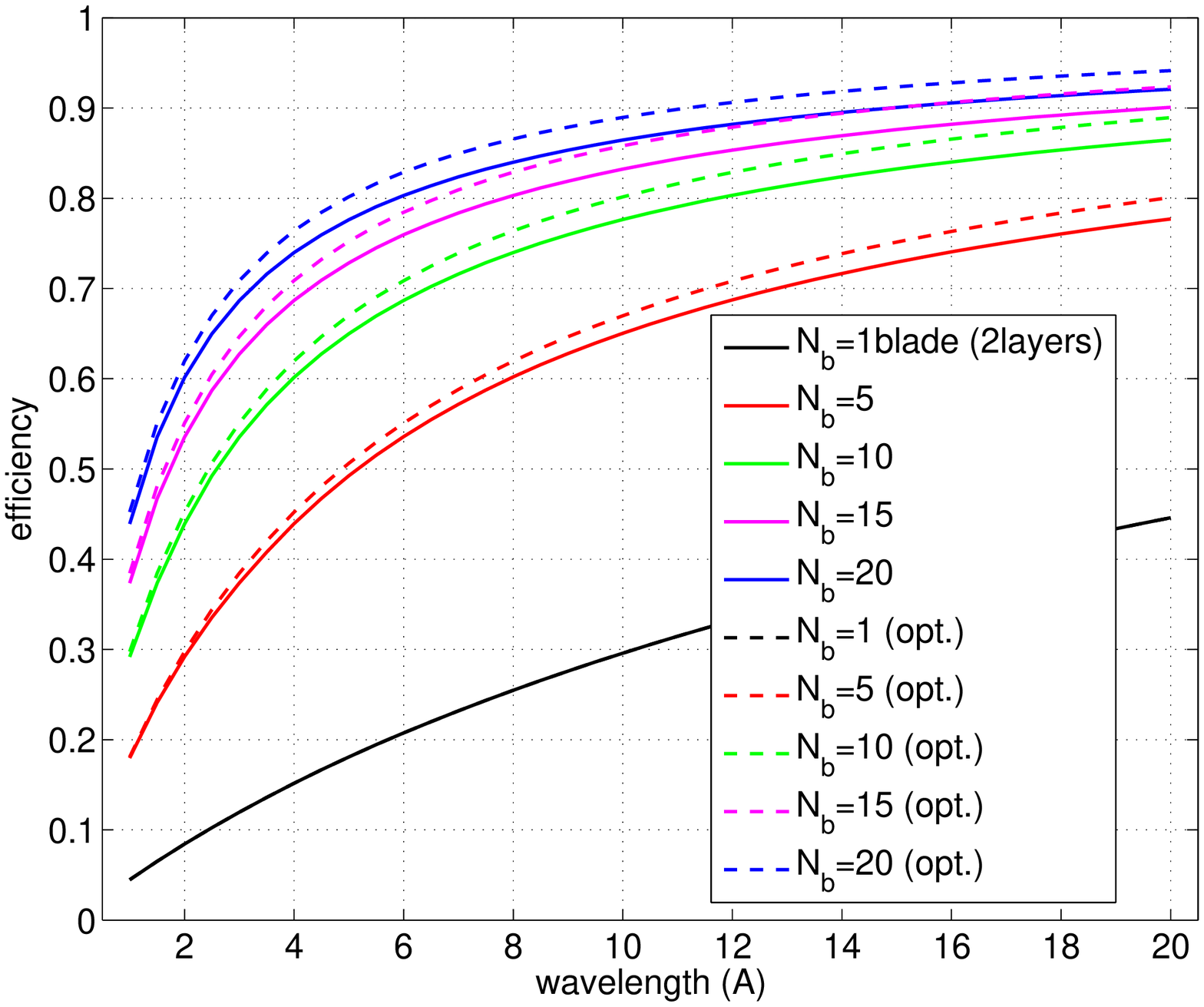}
\includegraphics[width=7.5cm]{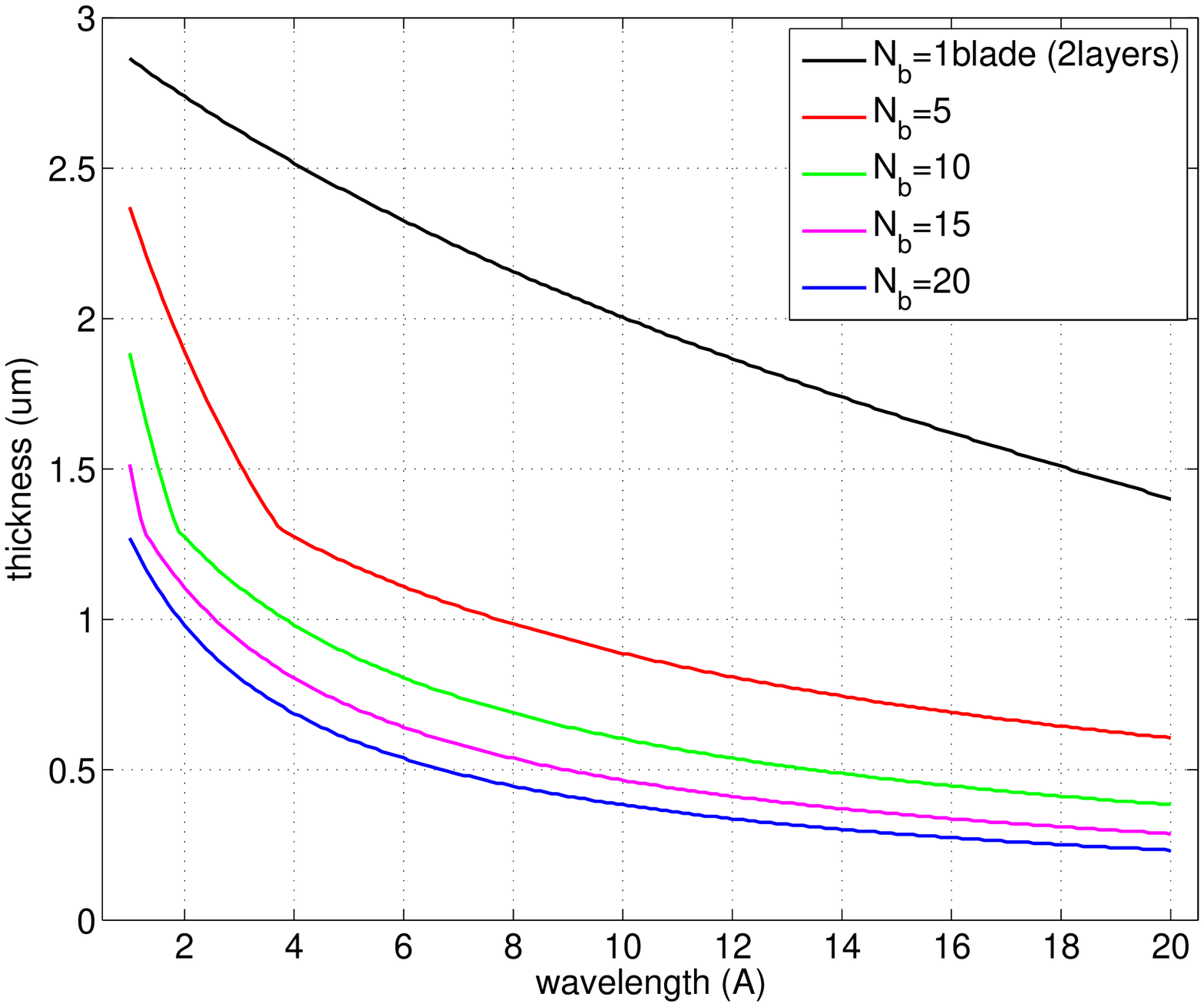}
\caption{\footnotesize Efficiency (left) and optimal thickness of
the identical blades (right) as a function of neutron wavelength for
a 2, 10, 20, 30 and 40 layers multi-layer detector. Solid lines
indicate the optimized efficiency, for each wavelength, for a
detector made up of blades of identical thicknesses; the dashed ones
indicate the monochromatic optimization using different
thicknesses.} \label{2optimizzd}
\end{figure}

We analyze a detector composed of $30$ successive converter layers
($15$ blades) crossed by the neutron beam at $90^{\circ}$ (like in
Figure \ref{multigridschet}). We consider $^{10}B_4C$ ($\rho=2.24\,
g/cm^3$) as converter; we neglect again the $6\%$ branching ratio of
$^{10}B$ neutron capture reaction. A $100\,KeV$ energy threshold is
applied and particle ranges turn out to be $R_1=3\, \mu m$
($\alpha$-particle) and $R_2 = 1.3\, \mu m$ ($^7Li$), for the $94\%$
branching ratio.
\\ Figures \ref{figMG30mon1p8} and \ref{figMG30mon20} show the
optimization result for this multi-layer detector; for a
monochromatic neutron beam of $1.8$\AA \, and $10$\AA. On the left,
the optimal thickness given by either Equations \ref{eqad11} or
\ref{eqad12} is plotted in red for each blade; for comparison we use
two similar detectors suitable for short and for long wavelengths in
which the blades are holding $1.2\, \mu m$ and $0.5\,\mu m$
thickness coating. Those values have been obtained by optimizing the
Equation \ref{eqad1sthick}, the efficiency for a detector holding
$15$ blades of all equal thicknesses for $1.8$\AA \, and for for
$10$\AA. The detector with $1.2\, \mu m$ coatings is very close to
the one presented in \cite{jonisorma}. On the right, in Figures
\ref{figMG30mon1p8} and \ref{figMG30mon20}, the efficiency
contribution of each blade is plotted, again for an optimized
detector for $1.8$\AA \, and for an optimization done for $10$\AA.
The expression of the efficiency as a function of the detector depth
is given by Equation \ref{eqad7} for each blade by fixing the index
$k$.
\begin{table}[!ht]
\caption{\footnotesize Efficiency for an optimized multi-layer
detector and for a detector which contains $15$ identical blades of
$1.2\, \mu m$ and $0.5\, \mu m$.} \centering
\begin{tabular}{|c|c|c|c|}
\hline \hline
wavelength (\AA)  & opt. detect. & $0.5\, \mu m$ detect.& $1.2\, \mu m$ detect. \\
\hline
1.8 &  0.525  & 0.388 & 0.510 \\
10  &  0.858  & 0.831 & 0.671 \\
\hline \hline
\end{tabular}
\label{tabeff23}
\end{table}
The whole detector efficiency is given in the end by summing all the
blades' contributions. The whole detector efficiency is displayed in
Table \ref{tabeff23} for the detector of Figures \ref{figMG30mon1p8}
and \ref{figMG30mon20}. By optimizing the detector for a given
neutron wavelength we gain only about $2\%$ efficiency which is
equivalent to add more layers to the detectors optimized to hold
identical blades.
\\ In Figure \ref{2optimizzd} is shown the
efficiency resulting from the monochromatic optimization process of
the individual blade coatings and the optimization for a detector
containing all identical blades (which thicknesses are shown on the
right for each neutron wavelength). Neutrons hit the layers at
$90^{\circ}$ and five cases have been taken into account with an
increasing number of layers. We notice that about for all neutron
wavelengths the gain in optimizing the detector with different
blades, let us to gain few percent in efficiency. The values in
Table \ref{tabeff23} are the values on the pink solid curve and the
dashed one at $1.8$\AA \, and at $10$\AA \, in Figure
\ref{2optimizzd}.
\\ Still referring to Figure \ref{2optimizzd}, we notice that a
detector with 15 individually optimized blades (30 layers) has about
the same efficiency (above $10$\AA) than a detector optimized to
contain 20 blades (40 layers) of equal thickness. On the other hand
for short wavelengths the difference is not very significant.
Moreover, there is also a trade off between the constraints of the
detector construction and the complexity of the blade production.

\subsection{Multi-layer detector optimization for a distribution of neutron wavelengths}\label{monomulti990}
In this case it is not possible to start the optimization from the
last blade because the thicknesses of the previous layers will
affect the neutron wavelength distribution reaching the deeper
laying blades. We have in this case to optimize an $N_b$-dimensional
function at once. Therefore, the $N_b$-dimensional equation
$\nabla\varepsilon_{tot}^w=0$ has to be solved:
\begin{equation}\label{eqad18}
\nabla_k\varepsilon_{tot}^w = \int_{0}^{+\infty}w\left(\lambda
\right) \frac{\partial \varepsilon_{tot}}{\partial d_{k}} \,
d\lambda=0
\end{equation}
$\frac{\partial \varepsilon_{tot}}{\partial d_{k}}$ is an expression
similar to Equation \ref{eqad16} provided that we impose
$d_{BSk}=d_{Tk} \, \forall k=1,2,\dots,N_b$.
\\ In order to optimize a detector for a given neutron
wavelength distribution $w\left(\lambda \right)$, the following
system of $N_b$ equations in $N_b$ unknown ($d_k$) has to be solved:
\begin{equation}\label{eqad19}
\begin{cases}
\int_{0}^{+\infty}w\left(\lambda \right)\left[ \frac{\partial
\varepsilon_1(d_k)}{\partial d_k}-2\Sigma \cdot
\sum_{p=\left(k+1\right)}^{N_b} \varepsilon_1(d_p) \cdot e^{-
2\left( \sum_{i=1}^{\left(p-1\right)} d_i \right)\cdot \Sigma
}\right]  \, d\lambda=0 &\mbox{if \,} k = 1 \\
\int_{0}^{+\infty}w\left(\lambda \right)\left[ \frac{\partial
\varepsilon_1(d_k)}{\partial d_k}\cdot e^{- 2\left(
\sum_{j=1}^{\left(N_b-1\right)} d_j \right)\cdot \Sigma
}\right.+\\-\left.2\Sigma \cdot \sum_{p=\left(k+1\right)}^{N_b}
\varepsilon_1(d_p) \cdot e^{- 2\left( \sum_{i=1}^{\left(p-1\right)}
d_i \right)\cdot \Sigma }\right] \, d\lambda=0 &\mbox{if \, }
1<k<N_b \\ \int_{0}^{+\infty}w\left(\lambda
\right)\left[\frac{\partial \varepsilon_1(d_k)}{\partial d_k}\cdot
e^{- 2\left( \sum_{j=1}^{\left(N_b-1\right)} d_j \right)\cdot \Sigma
}\right] \, d\lambda=0
 &\mbox{if \, } k = N_b
\end{cases}
\end{equation}
\\ We recall that $\varepsilon_1(d_k)$ and $\Sigma$ are function of
$\lambda$ and $\varepsilon_1(d_k)$ is the blade efficiency defined
in Equations \ref{eqac2} and \ref{eqac4}; its derivative
$\frac{\partial \varepsilon_1(d_k)}{\partial d_k}$ was already
calculated in the Equations \ref{eqac14} and \ref{eqac17} (Section
\ref{Sect2laysub}).
\\ The system of equations \ref{eqad19} can easily be solved numerically.

\subsubsection{Flat neutron wavelength distribution example}
\begin{figure}[!ht]
\centering
\includegraphics[width=7.5cm,angle=0,keepaspectratio]{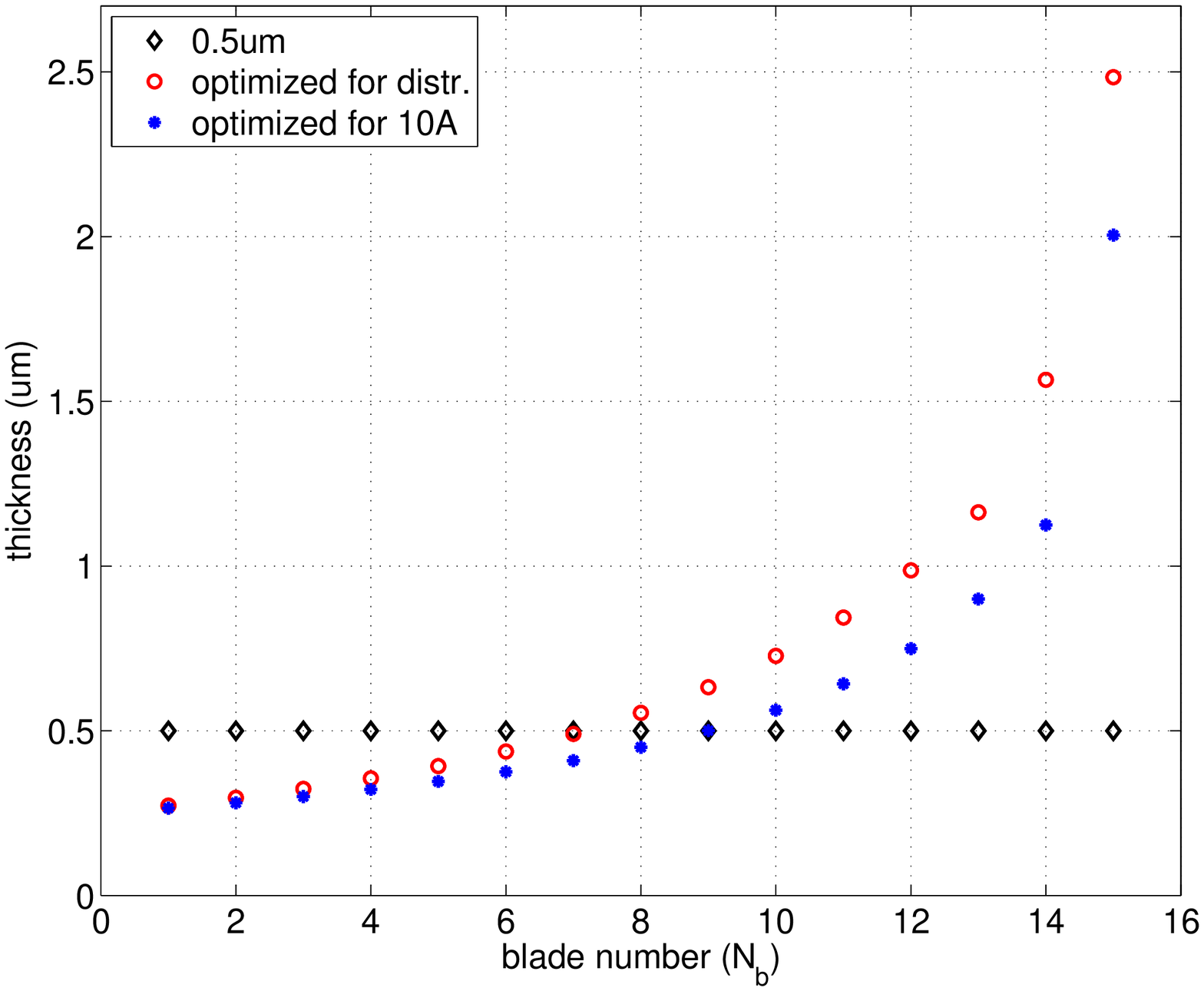}
\includegraphics[width=7.5cm,angle=0,keepaspectratio]{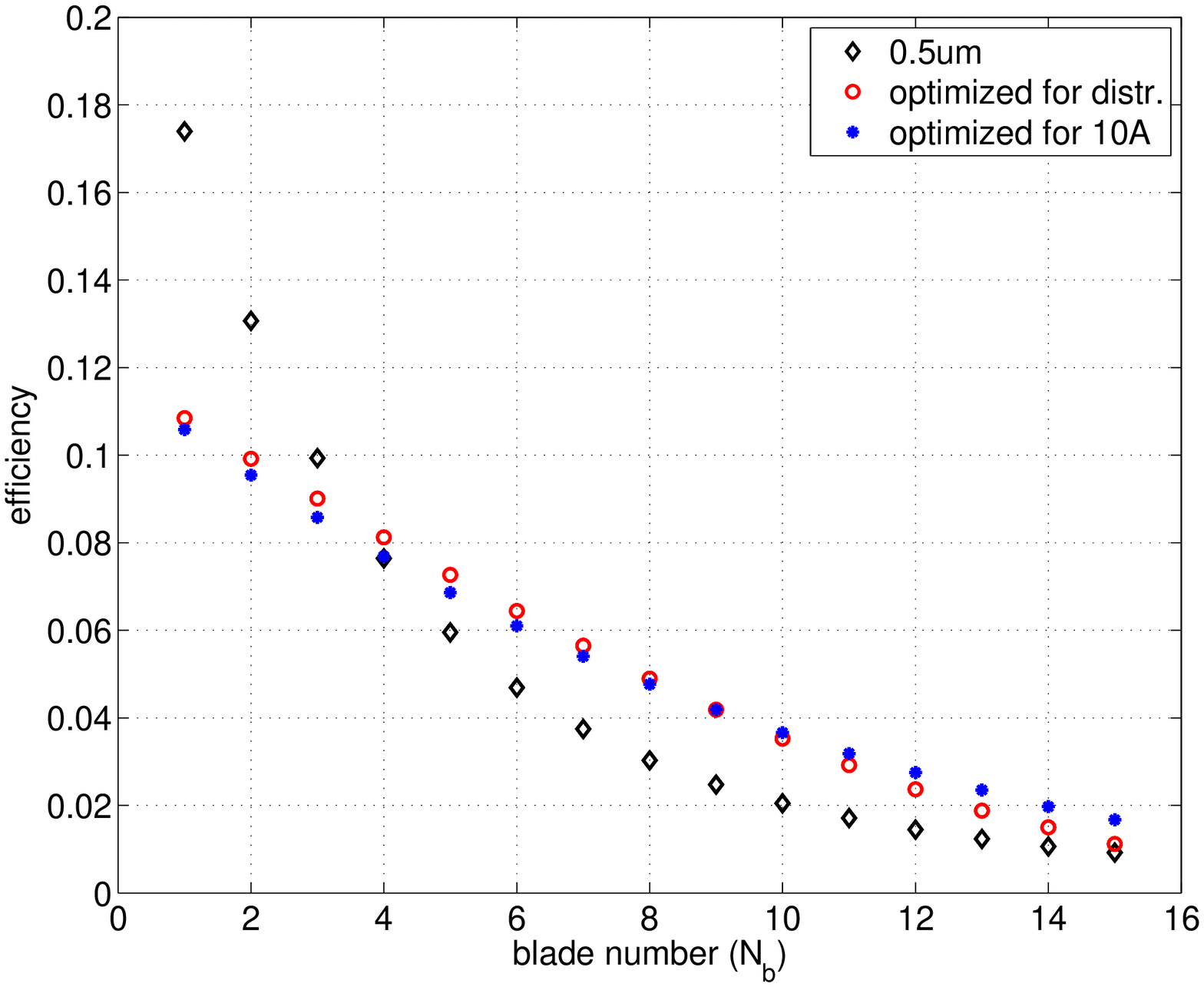}
 \caption{\footnotesize Thicknesses of the blade coatings (left) and their efficiency contribution (right),
 for a detector made up of $15$ blades of $0.5\, \mu m$, for
a detector optimized for the flat distribution
 of wavelengths and for a detector optimized for $10$\AA.} \label{figMG30flat102}
\end{figure}

We take a flat distribution
$w\left(\lambda\right)=\frac{1}{\lambda_2-\lambda_1}$ between the
two wavelengths $\lambda_1=1$\AA \, and $\lambda_2=20$\AA \, as in
Section \ref{Sect2laysub} for the single blade case. In Figure
\ref{figMG30flat102} the thicknesses of each of the blade coatings
and each blade efficiency contribution for a $30$-layer detector are
shown. Three detectors are compared, the one of simplest
construction is a detector holding $15$ identical blades of $0.5\,
\mu m$ coating thickness, the second is a detector optimized
according to Equation \ref{eqad19} for that specific flat
distribution and the last is a detector that has been optimized for
a single neutron wavelength of $10$\AA \, conforming to Equations
\ref{eqad11}, \ref{eqad12} and \ref{eqad13}. The fact to have a
contribution of wavelengths shorter than $10$\AA \, in the case of
the red line makes the coating thicknesses larger compared to the
blue curve.
\\ As a result, frontal layers are slightly more efficient for the
optimized detector than for the one optimized for $10$\AA; on the
other hand, deep layers lose efficiency.
\begin{figure}[!ht]
\centering
\includegraphics[width=7.5cm,angle=0,keepaspectratio]{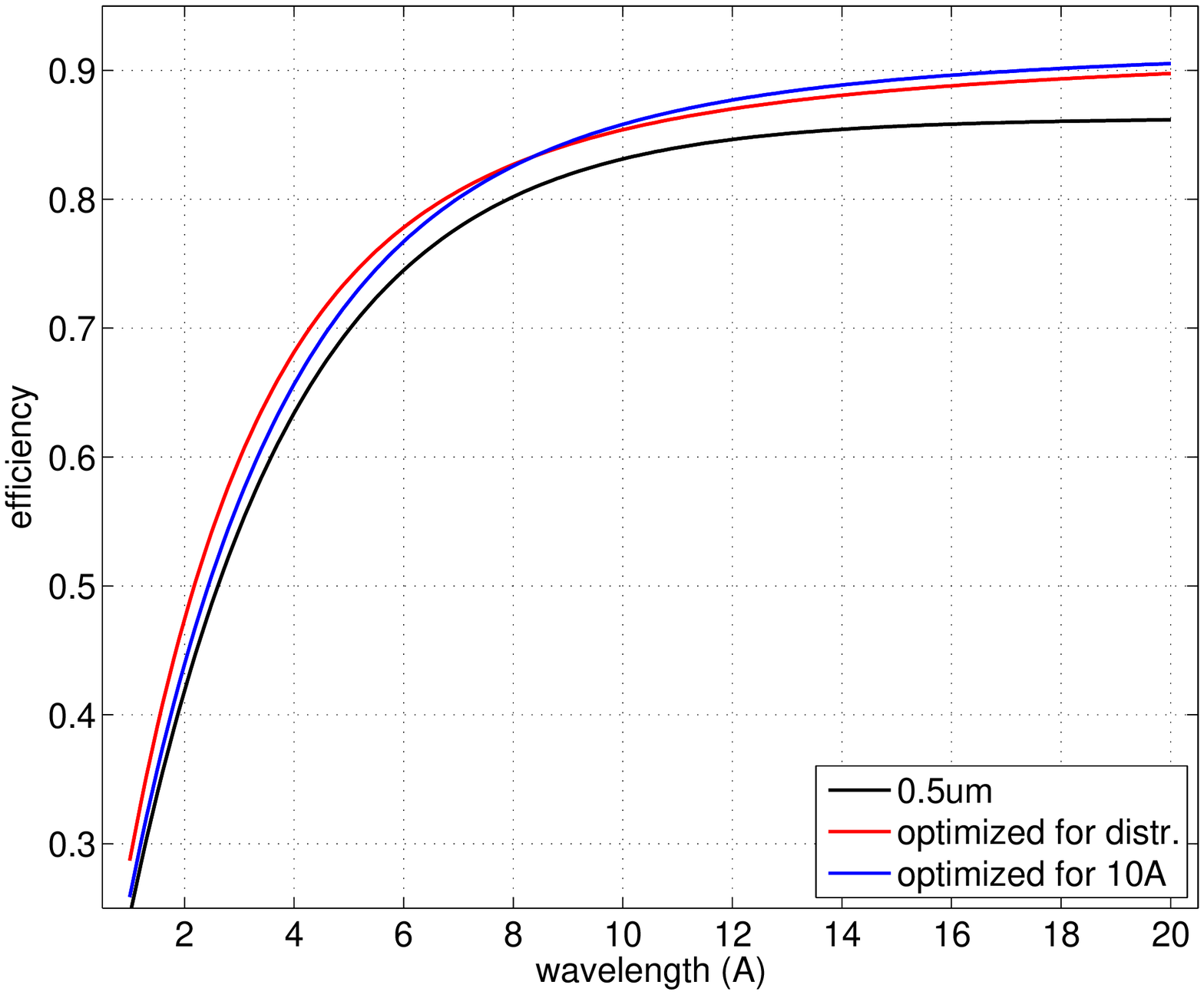}
\includegraphics[width=7.5cm,angle=0,keepaspectratio]{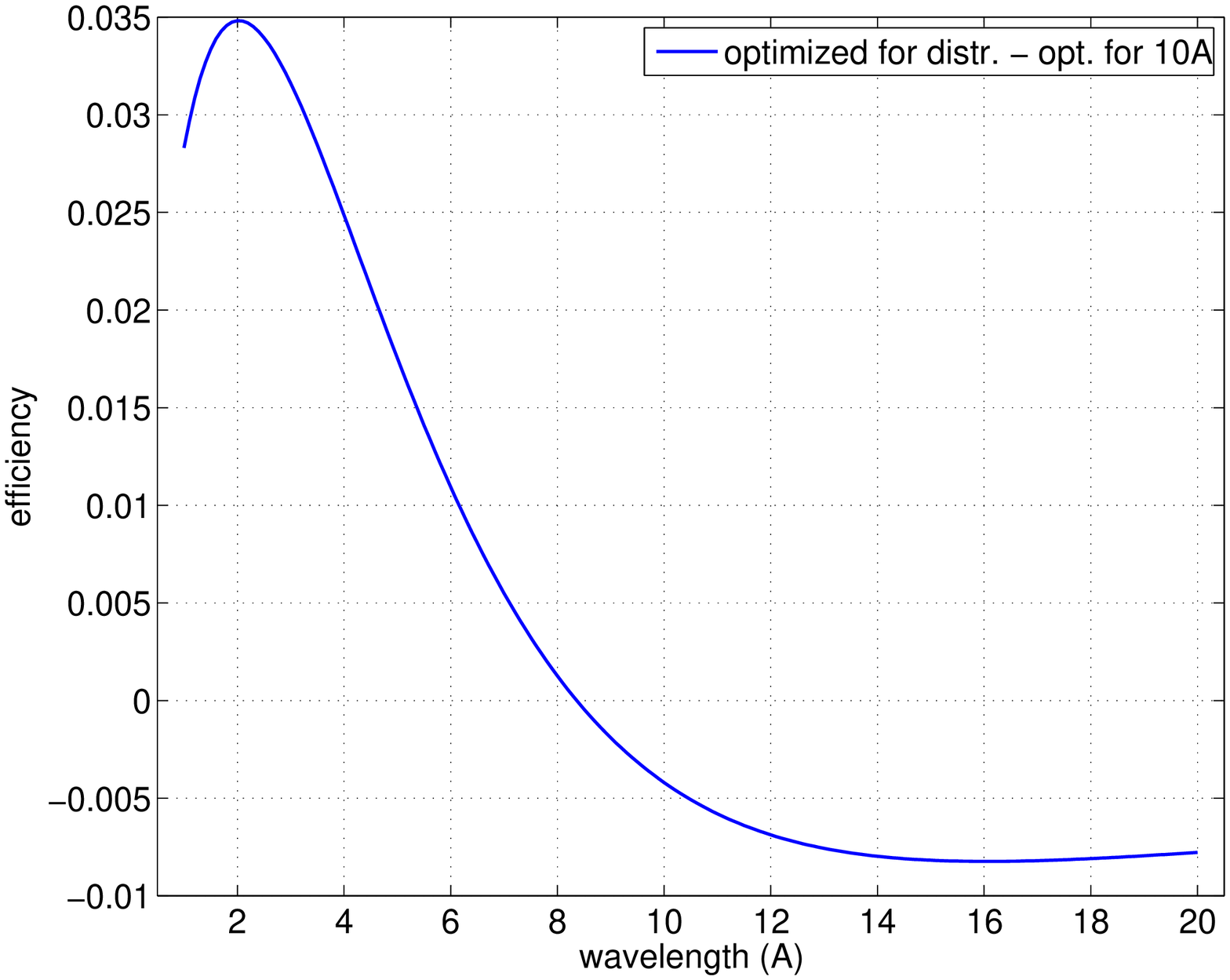}
 \caption{\footnotesize Efficiency as a function of neutron wavelength (left) for a detector made
 up of $15$ identical coating thickness blades of $0.5\, \mu m$, for a detector optimized for the flat distribution
 of wavelengths and for a detector optimized for $10$\AA. Difference between the efficiencies for a detector optimized
 for a flat distribution and for $10$\AA \, as a function of neutron wavelength (right).} \label{figMG30flat103}
\end{figure}
Figure \ref{figMG30flat103} shows the three detector efficiencies as
a function of neutron wavelength. By comparing red and blue lines,
of which the difference is plotted on the right plot, the optimized
detector gains efficiency on shorter wavelengths but loses on
longer. Moreover, we notice that the optimization process explained
in this section let to gain at most $3.5\%$ at short wavelengths
while losing less than $1\%$ on longer ones. The weighted efficiency
over $w\left(\lambda\right)$ is shown in Table \ref{tabeff569}.
\begin{table}[!ht]
\caption{\footnotesize Averaged efficiency over the flat neutron
wavelength distribution ($1$\AA-$20$\AA) for a detector which
contains $15$ identical blades of $0.5\, \mu m$, for an optimized
multi-layer detector for that specific flat distribution and for a
detector optimized for $10$\AA (Energy threshold of $100\,KeV$ is
applied).} \centering
\begin{tabular}{|c|c|c|}
\hline \hline
 opt. detect. & opt. detect. for 10\AA & $0.5\, \mu m$ detect. \\
\hline
0.796 & 0.793 & 0.764 \\
\hline \hline
\end{tabular}
\label{tabeff569}
\end{table}
We can conclude that if we are interested in optimizing a detector
in a given interval of wavelengths without any preference to any
specific neutron energy; optimizing according to Equation
\ref{eqad19} does not give a big improvement in the average
efficiency compared to optimizing for the neutron wavelength
distribution barycenter (about $10$\AA).
\\ Although the averaged efficiency for the optimized detector
in the neutron wavelength range differs from the one optimized for
$10$\AA \, only by $0.3\%$ one can be interested to have a better
efficiency for shorter wavelengths rather than for longer. It is in
this case that the optimization process can play a significant role.
On that purpose let's move to the following example.

\subsubsection{Hyperbolic neutron wavelength distribution example}
\begin{figure}[!ht]
\centering
\includegraphics[width=7.5cm,angle=0,keepaspectratio]{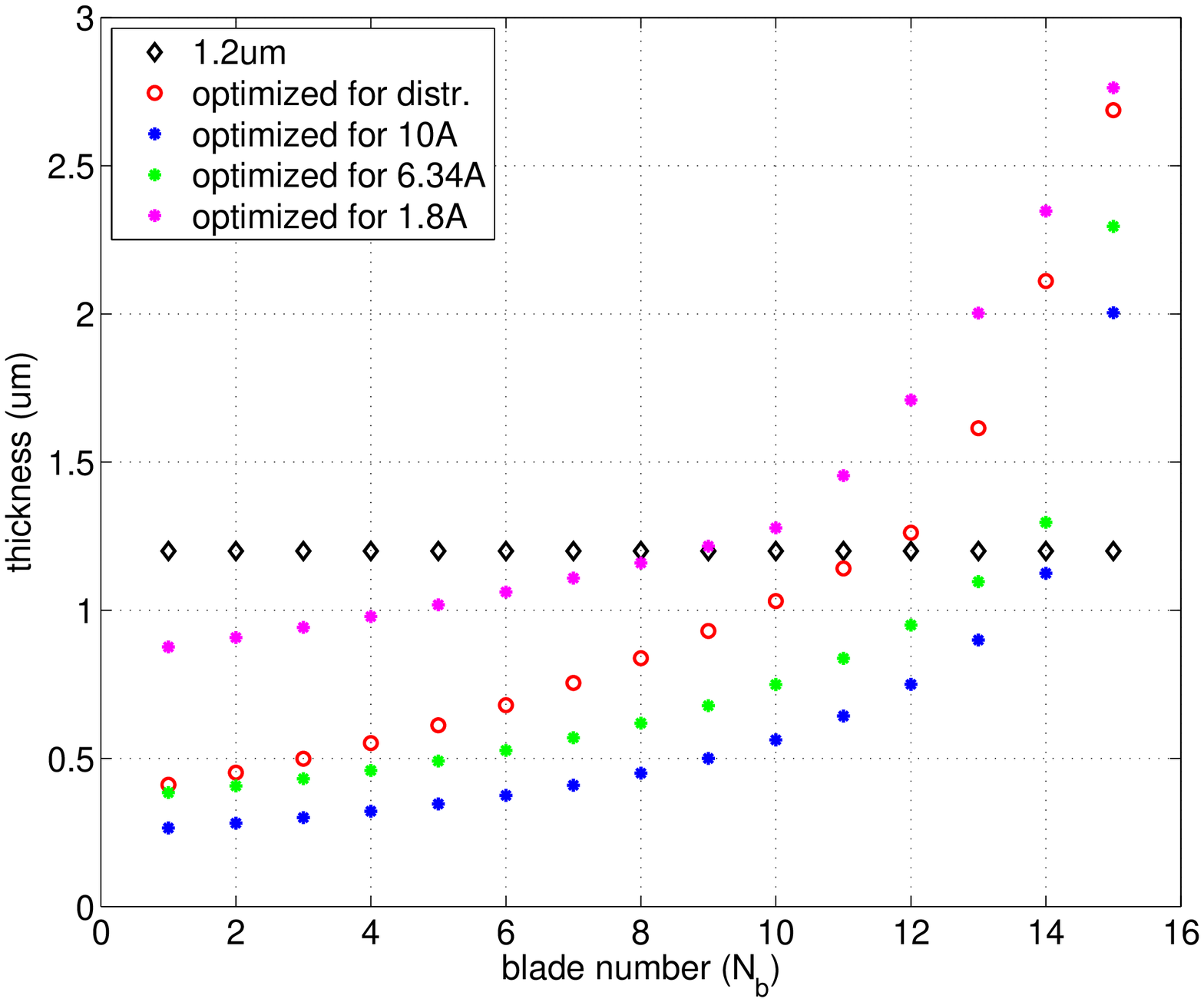}
\includegraphics[width=7.5cm,angle=0,keepaspectratio]{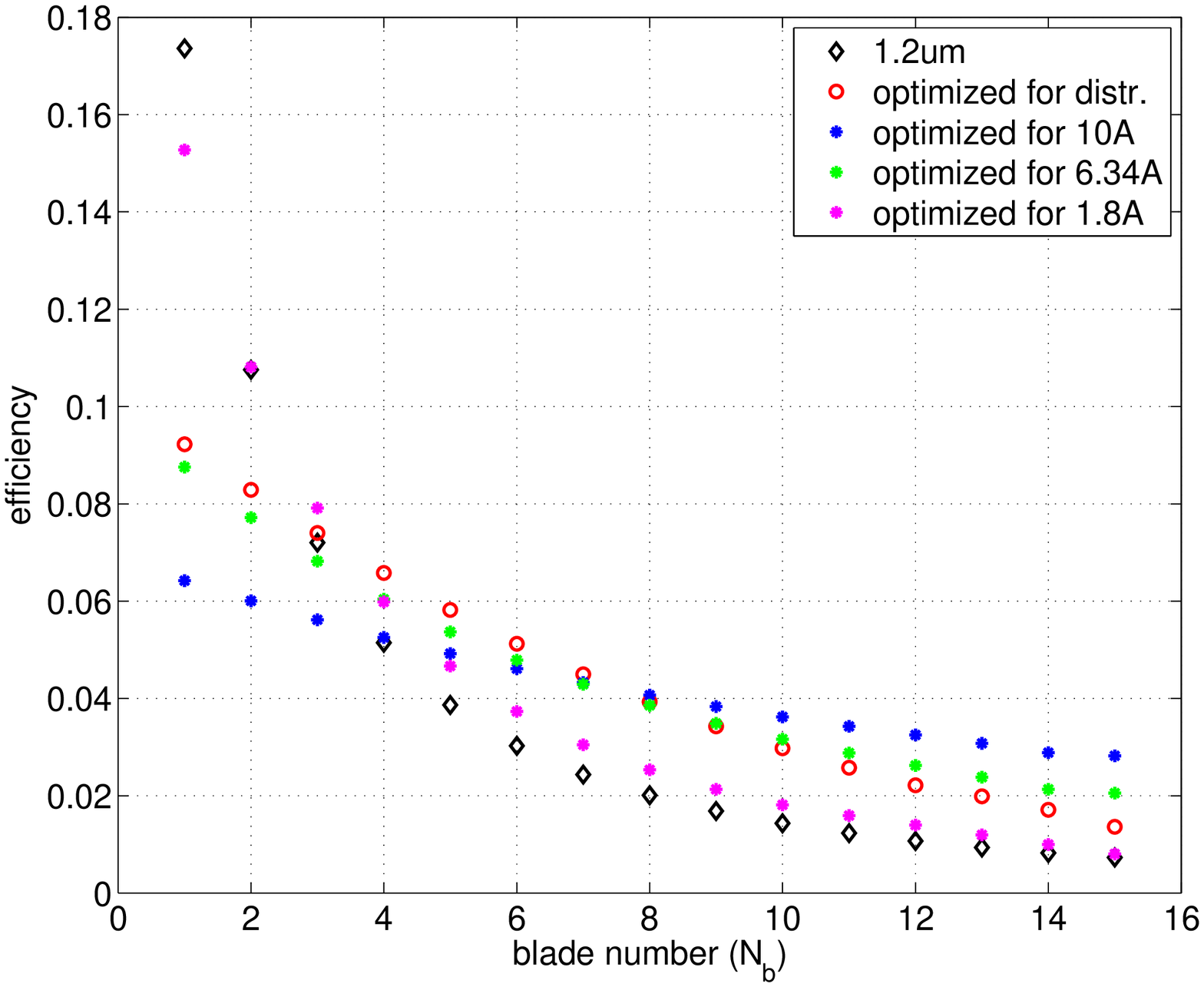}
 \caption{\footnotesize Thicknesses of the blades coatings (left) and their efficiency contribution (right), for a detector made
 up of $15$ identical coating thickness blades of $1.2\, \mu m$, for a detector optimized for an hyperbolic distribution
 of wavelengths and for a detector optimized for $10$\AA, $6.34$\AA \, and for for $1.8$\AA.}\label{figMG30iperb102}
\end{figure}

We consider a hyperbolic neutron wavelength distribution between
$\lambda_1=1$\AA \, and $\lambda_2=20$\AA.
\begin{equation}\label{eqad20}
w\left(\lambda\right)=\frac{1}{
\ln\left(\frac{\lambda_2}{\lambda_1}\right)} \cdot \frac{1}{\lambda}
\end{equation}
\\ This optimization aims to give equal importance to bins on a logarithmic
wavelength scale. The barycenter of the wavelength distribution
corresponds to
$\int_{\lambda_1}^{\lambda2}w\left(\lambda\right)\lambda \,
d\lambda=6.34$\AA.
\\ In Figure \ref{figMG30iperb102} are shown the
thicknesses of each blade coating and the efficiency as a function
of the depth direction in the detector for a $30$-layer detector.
Five detectors are compared, the one of $1.2\, \mu m$ coating
thickness, a detector optimized according to Equation \ref{eqad19}
for that specific hyperbolic distribution, a detector that has been
optimized for a single neutron wavelength of $10$\AA, $1.8$\AA, and
for the barycenter of the distribution.
\begin{figure}[!ht]
\centering
\includegraphics[width=7.5cm,angle=0,keepaspectratio]{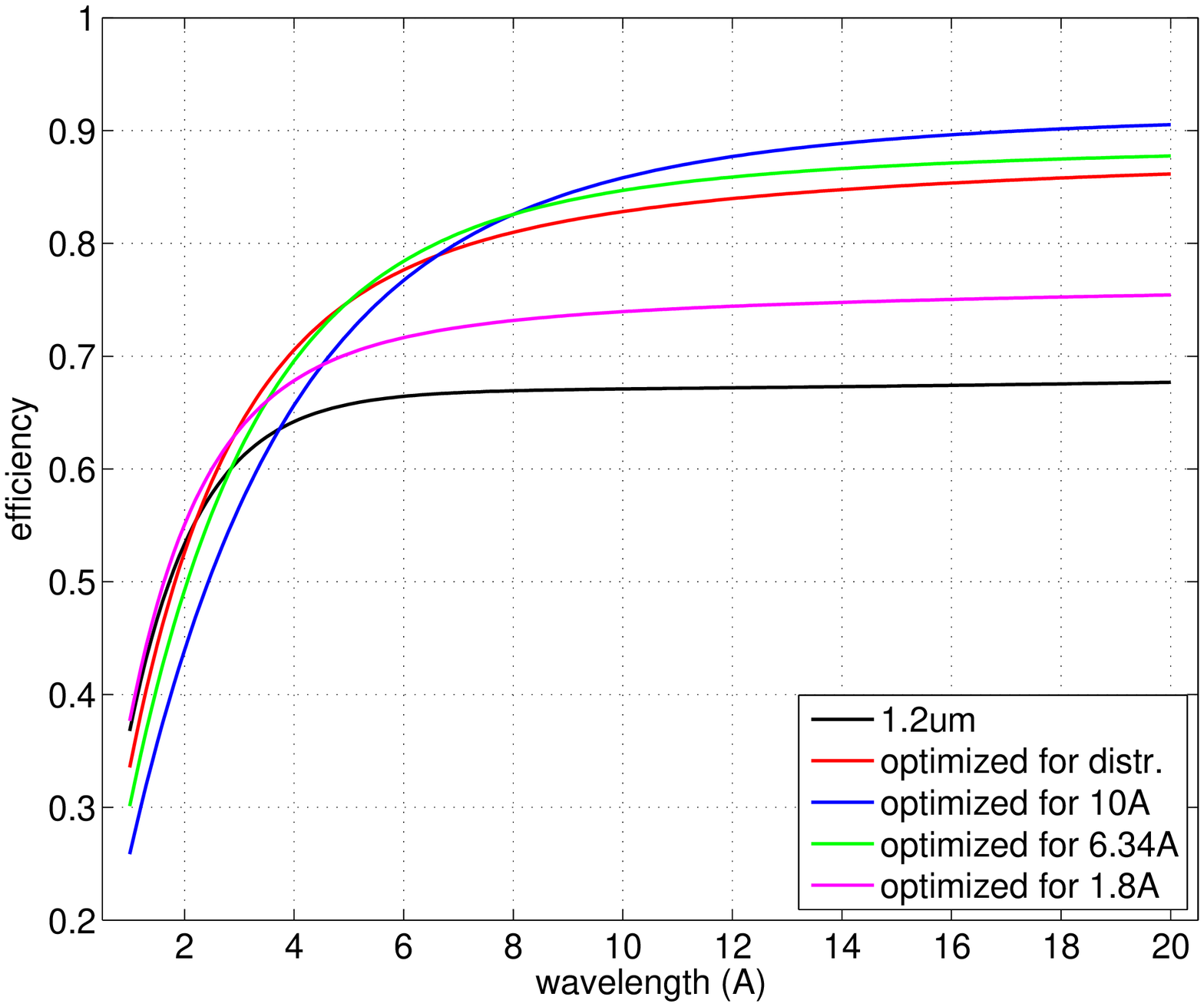}
\includegraphics[width=7.5cm,angle=0,keepaspectratio]{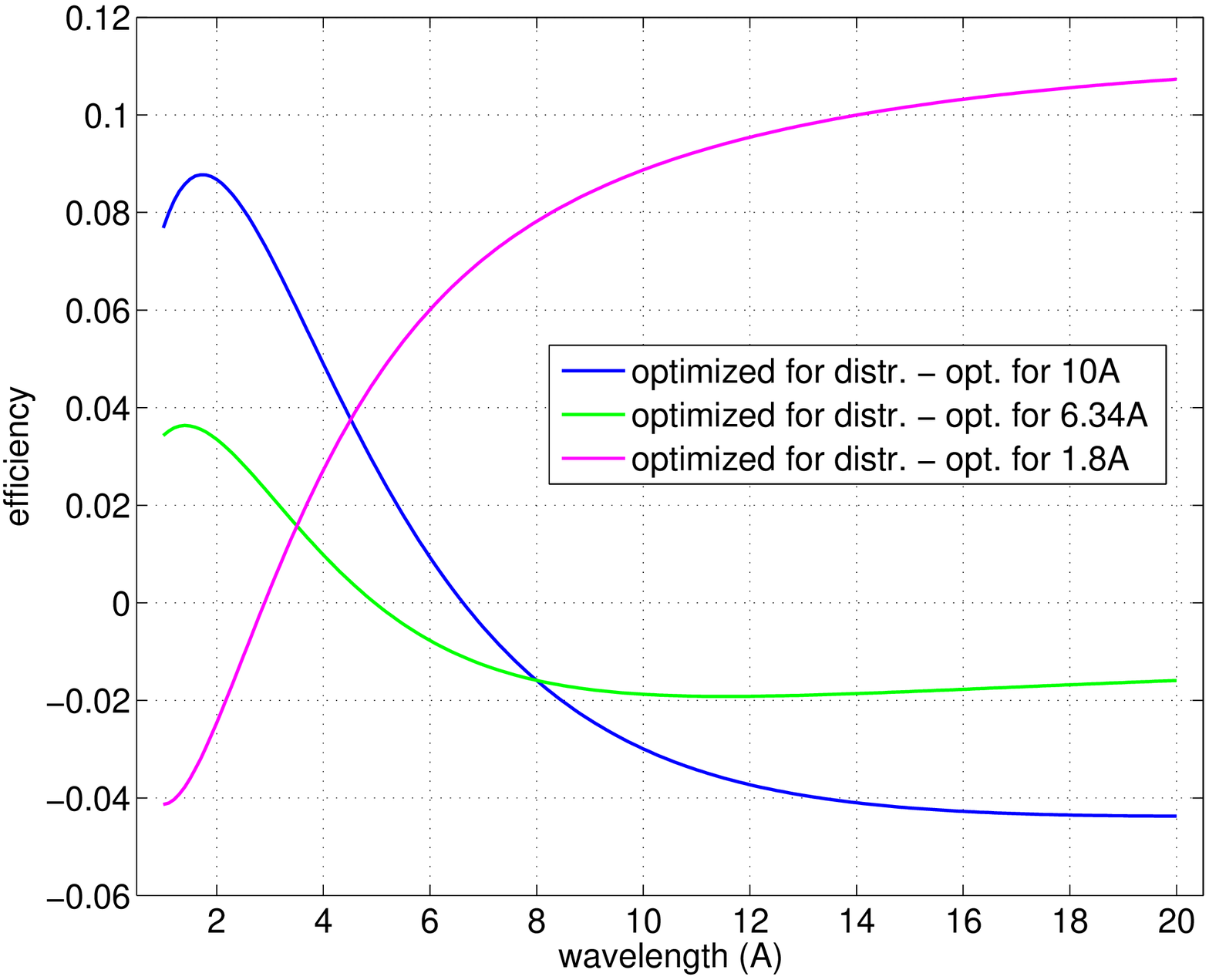}
 \caption{\footnotesize Efficiency as a function of neutron wavelength (left) for a detector made
 up of $15$ identical coating thickness blades of $1.2\, \mu m$, for a detector optimized for an hyperbolic distribution
 of wavelengths and for a detector optimized for $10$\AA \, and for $1.8$\AA. Difference between the efficiencies for a detector optimized
 for a flat distribution and for $10$\AA, $6.34$\AA \, and for $1.8$\AA \, as a function of neutron wavelength (right).}\label{figMG30iperb103}
\end{figure}
\begin{table}[!ht]
\caption{\footnotesize Averaged efficiency over the hyperbolic
distribution for a detector which contains $15$ identical blades of
$1.2\, \mu m$, for an optimized multi-layer detector for that
specific distribution and for a detector optimized for $10$\AA,
$6.34$\AA \, and for $1.8$\AA \, (Energy threshold of $100\,KeV$
applied).} \centering
\begin{tabular}{|c|c|c|c|c|}
\hline \hline
opt. detect. & opt. $10$\AA & opt. $6.34$\AA  & opt. $1.8$\AA & $1.2\, \mu m$ detect.\\
\hline
0.671 & 0.641 & 0.664 & 0.639 & 0.597 \\
\hline \hline
\end{tabular}
\label{tabeff570}
\end{table}
\\ Figure \ref{figMG30iperb103} shows the five detector efficiencies as a
function of wavelength and their difference on the right plot. By
comparing the red (optimized detector) and the blue (detector
optimized for  $10$\AA) lines, of which the difference is plotted in
blue on the right plot, we notice that the detector optimized for
such a distribution gains about $9\%$ efficiency at short
wavelengths and loses about $4\%$ at high wavelengths. A detector
conceived for short wavelengths, such as the one represented by the
pink line, has an opposite behavior instead. The distribution
optimized detector gains efficiency for long wavelengths reaching
about $11\%$. Moreover, a detector optimized for the barycenter of
the neutron wavelength distribution, instead does not differ more
than about $4\%$ over the whole wavelength interval, as in the case
of a uniform distribution. By only comparing the averaged
efficiencies, shown in Table \ref{tabeff570}, it seems that there is
not a big improvement in the detector efficiency which is only about
$3\%$ for both $10$\AA \, and $1.8$\AA \, optimized detectors with
respect to the distribution optimized detector. On the other hand,
the optimization procedure, explained in this section, shows that it
can lead to a significant efficiency improvement in certain neutron
wavelength ranges. Furthermore, as in the case of a flat
distribution, a detector optimized for a distribution according to
Equations \ref{eqad19}, does not show significant improvement in
performances with respect to a detector just optimized for its
barycenter.

\section{Considerations on solid converter Pulse Height Spectra}\label{SectThPHSCalc}
The physical model taken into account in \cite{gregor} and in
\cite{salvat} can be used as well to derive the analytical formula
for the Pulse Height Spectra (PHS). A similar work was done in
\cite{salvat} (see Appendix \ref{app2}) where only Monte Carlo
solutions were shown; here we want to use analytic methods to
understand the structure of the PHS.
\\ We make approximation mentioned in the introduction and we assume
either a simplified stopping power function (see Section
\ref{strapp33}) or one simulated with SRIM \cite{sri} for the
neutron capture fragments.
\\ We calculate the probability for a particle emitted from the conversion point to
travel exactly a distance $L$ on a straight line towards the escape
surface (see Figure \ref{coorsys}). This distance $L$ is related to
the charged particle remaining energy through the primitive function
of the stopping power. We will demonstrate that even under strong
approximations of the stopping power function the model still
predicts quite well the important physical features of the PHS.

\subsection{Back-scattering mode}\label{backscatt678}
The probability for a neutron to be captured at depth
$\left(x,x+dx\right)$ in the converter layer and for the capture
reaction fragment (emitted isotropically in $4\pi\,sr$) to be
emitted with an angle $\varphi = \arccos(u)$ (between
$\left(u,u+du\right)$) is:
\begin{equation}\label{eqae1}
p(x,u)dx \, du=\begin{cases} \frac 1 2 \cdot \Sigma e^{-\Sigma \cdot x} \, dx \, du &\mbox{if  \,} x\leq d\\
0 &\mbox{if \, } x > d
\end{cases}
\end{equation}
\\ Where $\Sigma$ is the macroscopic cross-section already defined in
Equation \ref{eqac2abc} and $d$ is the layer thickness. \\The
fragment will travel a distance $L$ across the converter layer if
$L=\frac{x}{u}$ and $u \in [0,1]$.
\\ The probability for a particle to travel a distance $\left(L,L+dL\right)$ across
the layer is then given by:
\begin{equation}\label{eqae3}
P(L)\,dL= \int_0^d \, dx \int_0^1 \, du\,\,
\delta\left(\frac{x}{u}-L\right) p(x,u) =\begin{cases} \frac{1}{2
L^2} \left( \frac{1}{\Sigma}-(\frac{1}{\Sigma}+L)e^{-\Sigma \cdot
L}\right)\,dL &\mbox{if  \,} L \leq d\\
\frac{1}{2 L^2} \left(
\frac{1}{\Sigma}-(\frac{1}{\Sigma}+d)e^{-\Sigma \cdot d}\right)\,dL
&\mbox{if \, } L > d
\end{cases}
\end{equation}
\\ It is sufficient to replace $\Sigma$ with $\frac{\Sigma}{\sin(\theta)}$ if neutrons
hit the layer under the angle $\theta$ with respect to the surface
(see Figure \ref{coorsys}). The demonstration is equivalent to the
one that can found in \cite{gregor} for the efficiency function, in
the PHS calculation $p(x,u)$ has to be changed as follows:
\begin{equation}\label{eqae4}
p(x,u,\theta)dx \, du=\begin{cases} \frac 1 2 \cdot \Sigma e^{-\Sigma \cdot \frac{x}{\sin{(\theta)}}} \frac{dx}{\sin{(\theta)}} \, du &\mbox{if  \,} x\leq d\\
0 &\mbox{if \, } x > d
\end{cases}
\end{equation}
\\ If $E(L)$ is the remaining energy of a particle that has traveled a distance
$L$ into the layer, $\frac{dE(L)}{dL}$ is the stopping power or
equivalently the Jacobian of the coordinate transformation between
$L$ and $E$.
\\ Once $P(L)dL$ is known we can calculate $Q(E)dE$, therefore:
\begin{equation}\label{eqae6}
Q(E)dE= P(L(E))\cdot \frac{1}{\left|\frac{dE}{dL}\right|} \cdot dE
\end{equation}
\\where $Q(E)dE$ is the probability that an incident neutron will give
rise to a release of an energy $\left(E,E+dE\right)$ in the gas
volume; hence it is the analytical expression for the PHS.

\subsubsection{PHS calculation using SRIM output files for Stopping Power}\label{SRIMsp33}
\begin{figure}[!ht]
\centering
\includegraphics[width=10cm]{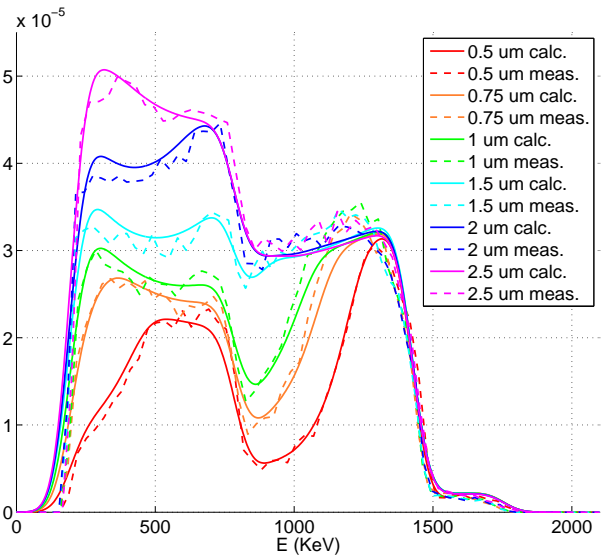}
\caption{\footnotesize Comparison between a PHS calculated and one
measured at ILL-CT2 on a $2.5$\AA \, neutron beam using a Multi-Grid
like detector \cite{jonisorma} where were mounted blades of
different thicknesses.}\label{phscfrMeas}
\end{figure}

We take the case of the $^{10}B$ reaction as example, however
results can be applied to any solid neutron converter. We recall the
energies carried for the $94\%$ branching ratio is $E_0=1470KeV$ for
the $\alpha$-particle and $E_0=830KeV$ for the $^7Li$; for the $6\%$
branching ratio, $E_0=1770KeV$ for the $\alpha$-particle and
$E_0=1010KeV$ for the $^7Li$. Referring to Equation \ref{eqae6}, the
stopping power $\frac{dE}{dL}$ used here was simulated with SRIM
\cite{sri} (see Figure \ref{stpowapprox45}) and $L(E)$ obtained by
numerical inversion of the stopping power primitive function, i.e.
the remaining energy inverse function.
\\ The complete PHS can be obtained
adding the four PHS in the case of $^{10}B$ according to the
branching ratio probability:
\begin{equation}\label{eqae7}
Q_{tot}(E)dE = \left(0.94 \cdot
\left(Q_{\alpha}(E)+Q_{^7Li}(E)\right)\right.+0.06 \cdot
\left.\left(Q_{\alpha}(E)+Q_{^7Li}(E)\right)\right)\cdot dE
\end{equation}
\\ Consequently the efficiency for a single layer can be calculated
by:
\begin{equation}\label{eqae8}
\varepsilon\left(E_{Th}\right)= \int_{E_{Th}}^{+\infty} Q_{tot}(E)dE
\end{equation}
\\ Where $E_{Th}$ is the energy threshold applied to cut the PHS. This
result is fully in agreement with what can be calculated by using
the Equations in \cite{gregor}.
\\ A Multi-Grid-like detector
\cite{jonisorma} was used to collect the data at ILL-CT2 where a
monochromatic neutron beam of $2.5$\AA \, is available. This
particular detector has the peculiarity that in each of its frames
blades of different coating thickness were mounted; as a result the
simultaneous PHS measurement for different layer thicknesses has
been possible. The blades are made up of an Aluminium substrate of
$0.5\,mm$ thickness coated on both sides by an enriched $^{10}B_4C$
layer \cite{carina}. Thicknesses available in the detector were:
$0.50\,\mu m$, $0.75\,\mu m$, $1\,\mu m$, $1.5\,\mu m$, $2\,\mu m$
and $2.5\,\mu m$.
\\ In our calculation we are not taking into account several processes,
such as wall effects, gas amplification and fluctuations, space
charge effects, etc. but only the neutron conversion and the
fragment escape. Moreover, while the calculation has an infinite
energy precision, this is not the case on a direct measurement
because many processes give a finite energy resolution.
\\ In order to be able to compare calculations and measurements,
after the PHS were calculated for the thicknesses listed above, we
convolve them with a gaussian filter of $\sigma=10\,KeV$. The
measured PHS were normalized to the maximum energy yield
($1770\,KeV$). An energy threshold of $180\,KeV$ was applied to the
calculation to cut the spectrum at low energies at the same level
the measured PHS was collected.
\\ We compare calculated and measured PHS in Figure \ref{phscfrMeas}; we can
conclude that the model gives realistic results in sufficient
agreement with the experimental ones, to be able to describe its
main features.

\subsubsection{PHS calculation using a strong approximation}\label{strapp33}
\begin{figure}[!ht]
\centering
\includegraphics[width=7.5cm,angle=0,keepaspectratio]{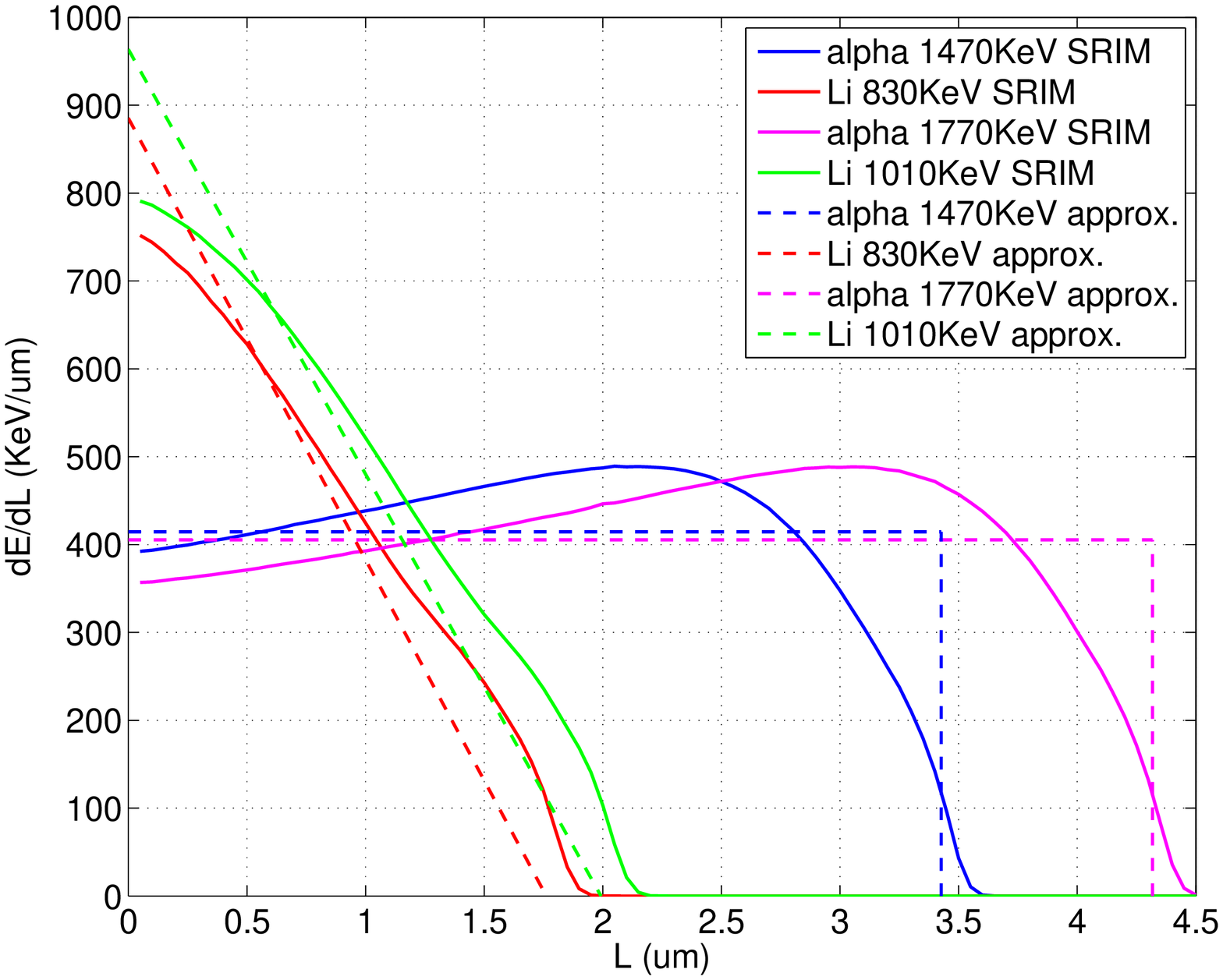}
\includegraphics[width=7.5cm,angle=0,keepaspectratio]{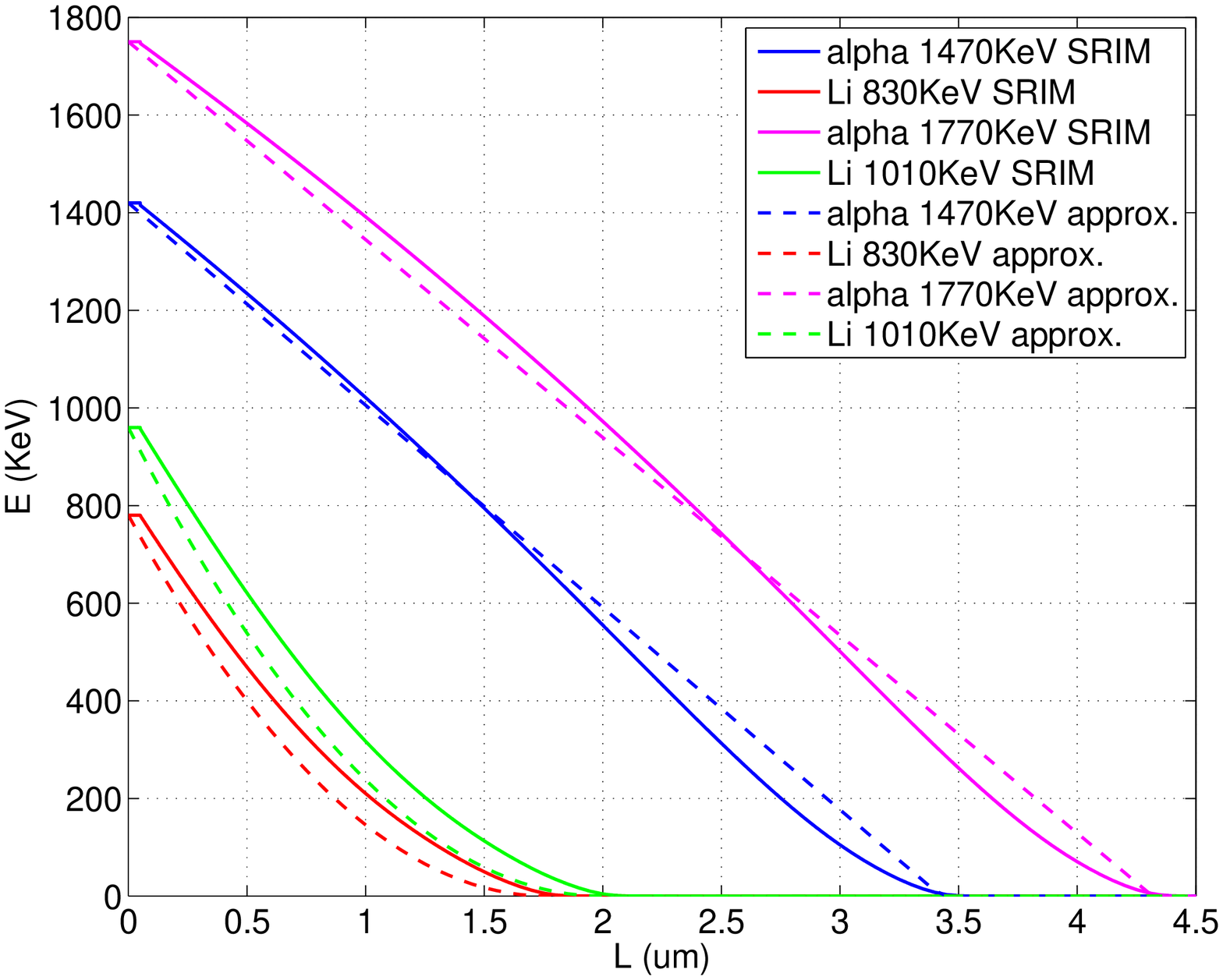}
\caption{\footnotesize Stopping power $\frac{dE}{dL}$ and its
primitive function $E(L)$ for $^{10}B$-reaction fragments, solid
curves are the functions obtained from SRIM \cite{sri}, dashed lines
are their approximated behaviors.} \label{stpowapprox45}
\end{figure}

\begin{figure}[!ht]
\centering
\includegraphics[width=7.5cm,angle=0,keepaspectratio]{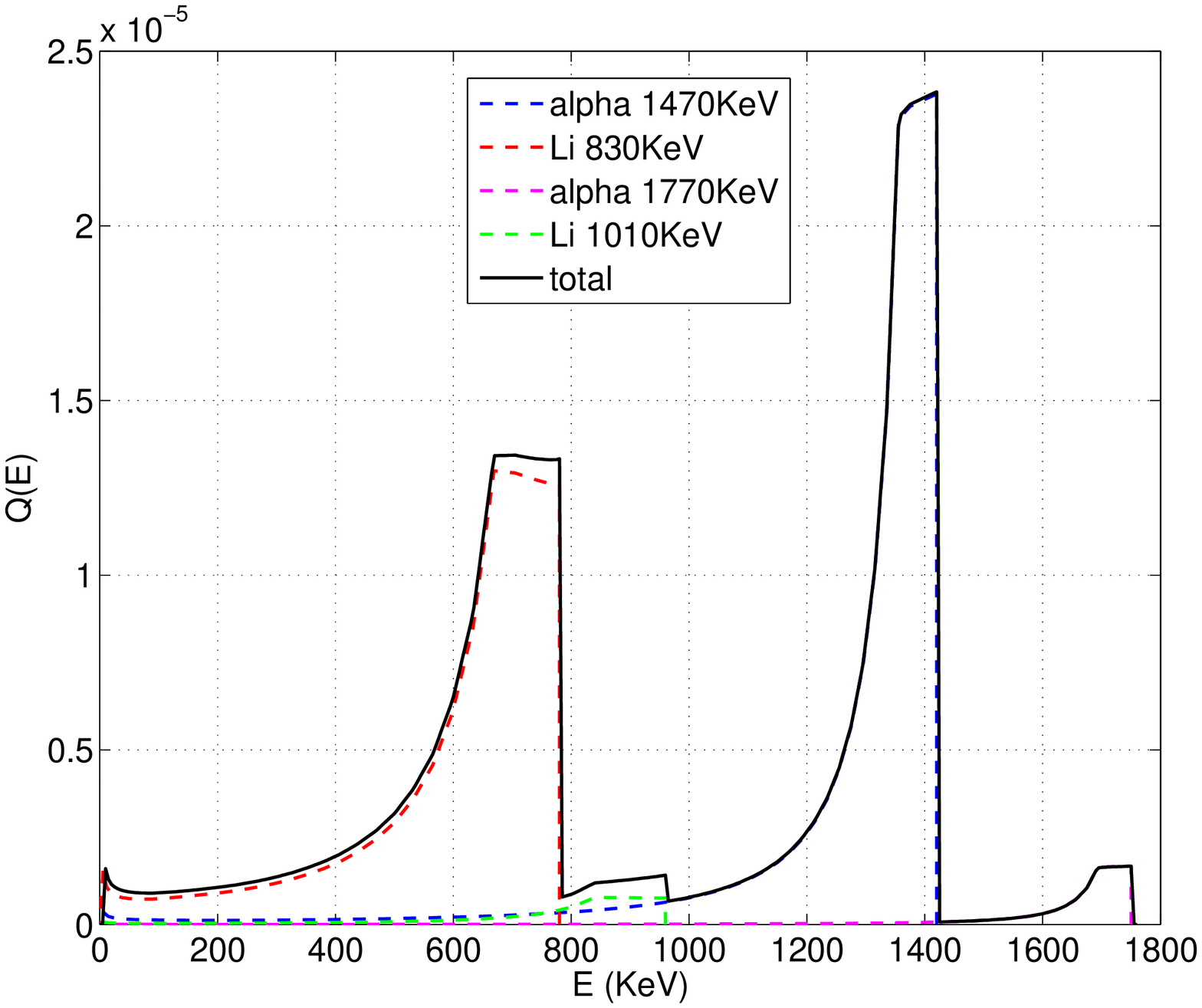}
\includegraphics[width=7.5cm,angle=0,keepaspectratio]{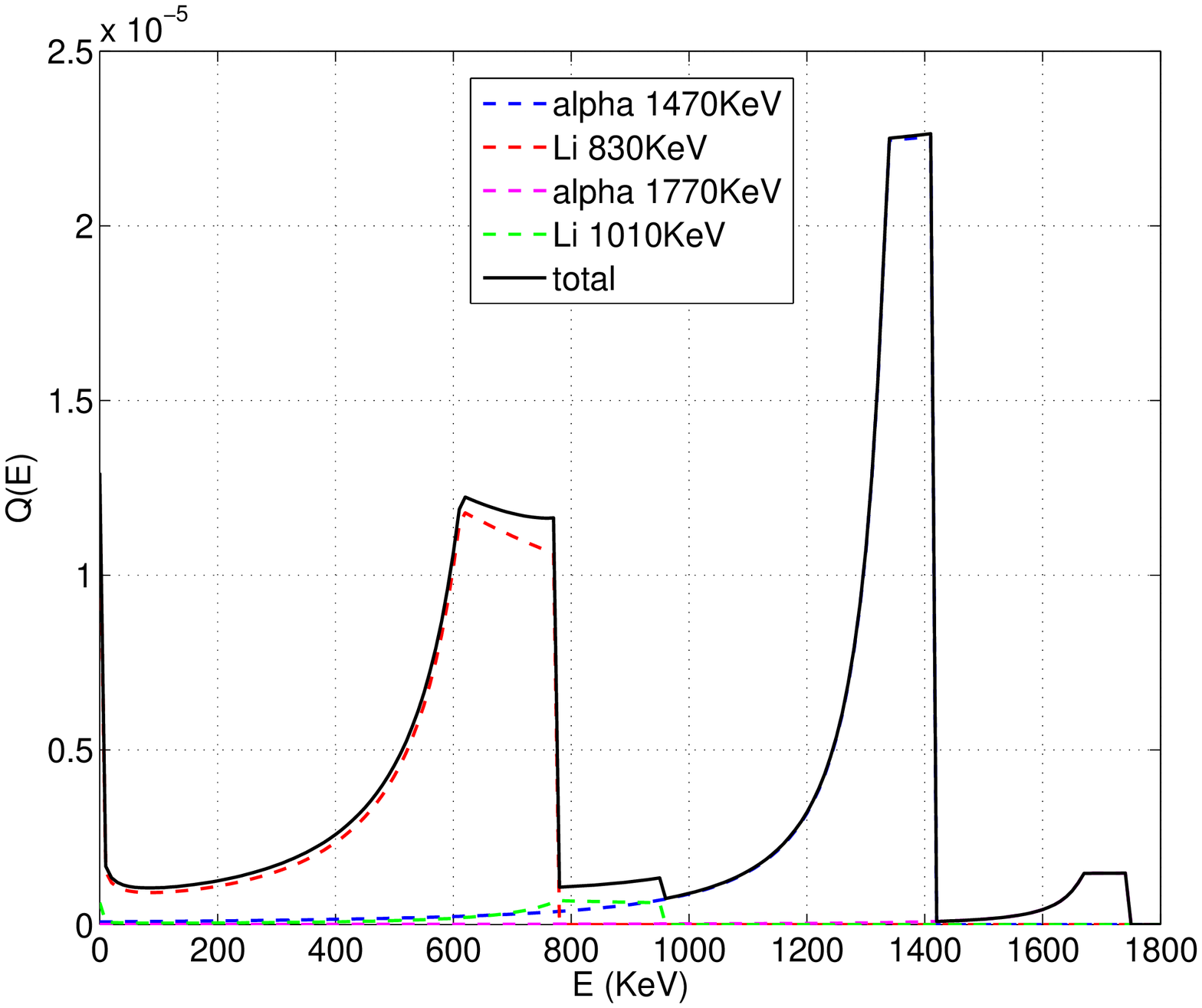}
 \caption{\footnotesize Calculated PHS using SRIM (left) and
 approximated (right) stopping power functions for a single back-scattering layer of $0.2\, \mu m$
 for $1.8$\AA \, and $90^{\circ}$ incidence.} \label{figPHSST401}
\end{figure}

\begin{figure}[!ht]
\centering
\includegraphics[width=7.5cm,angle=0,keepaspectratio]{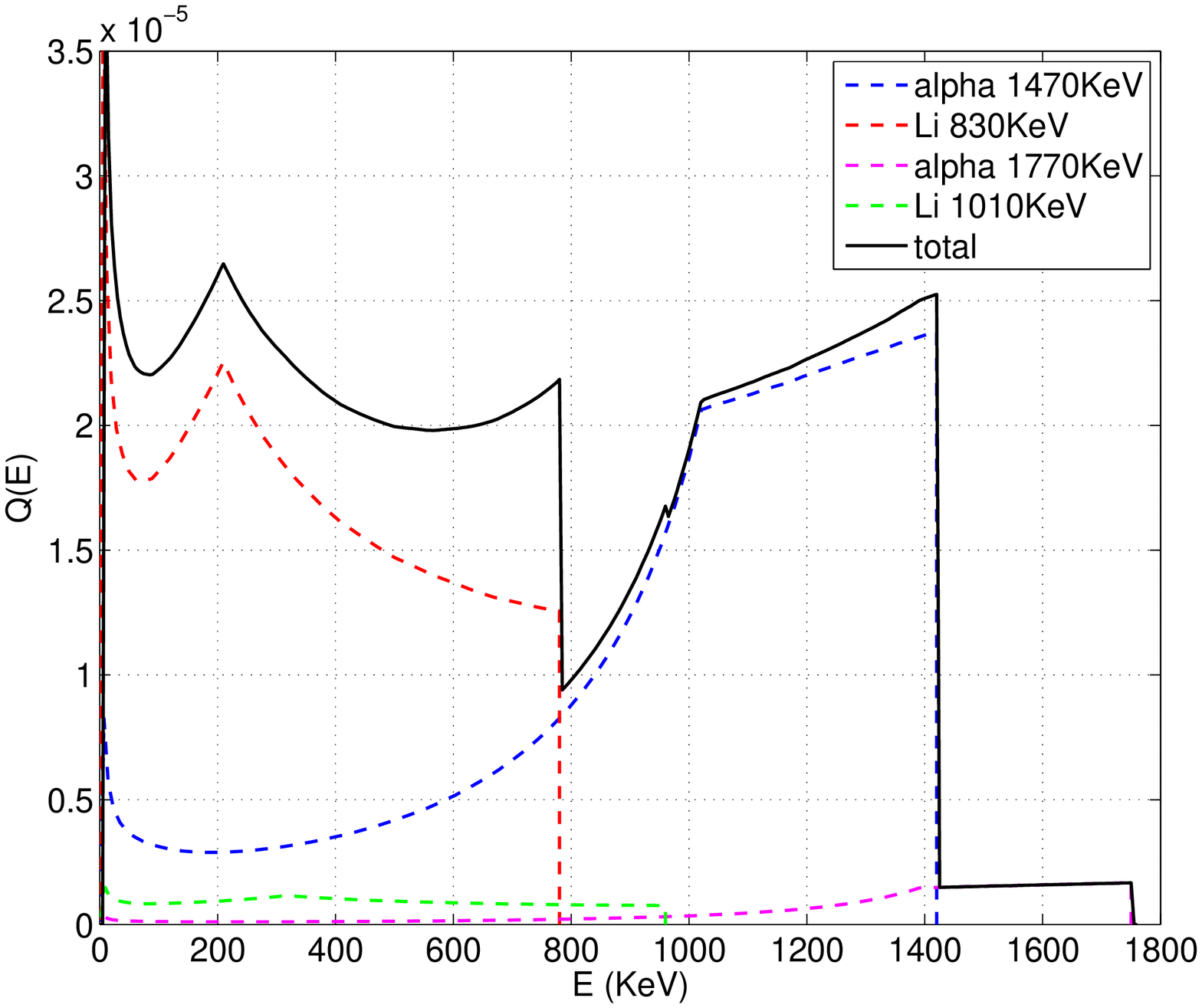}
\includegraphics[width=7.5cm,angle=0,keepaspectratio]{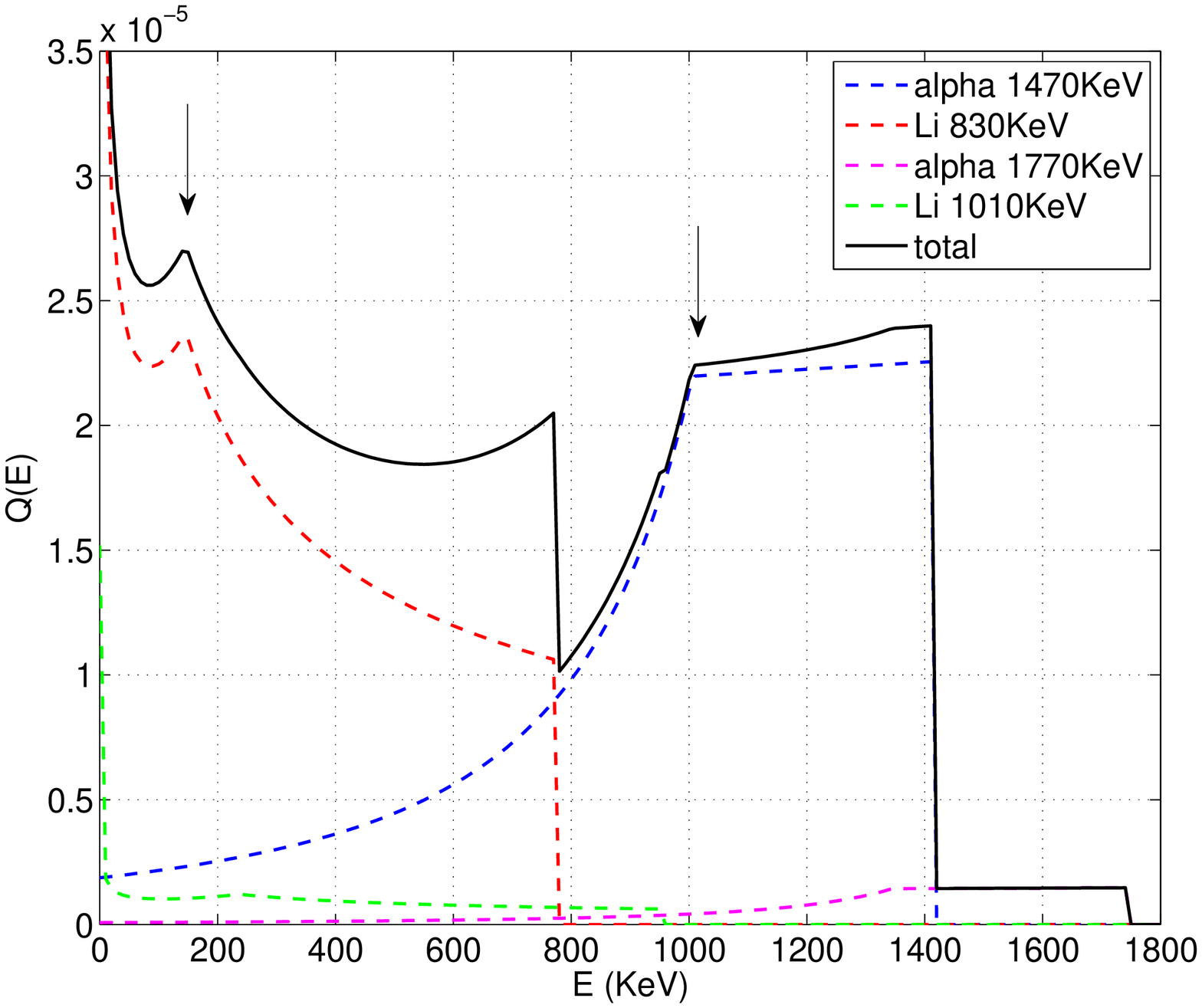}
 \caption{\footnotesize Calculated PHS using SRIM (left) and
 approximated (right) stopping power functions for a single back-scattering layer of $1\, \mu m$
 for $1.8$\AA \, and $90^{\circ}$ incidence.} \label{figPHSST402}
\end{figure}

\begin{figure}[!ht]
\centering
\includegraphics[width=7.5cm,angle=0,keepaspectratio]{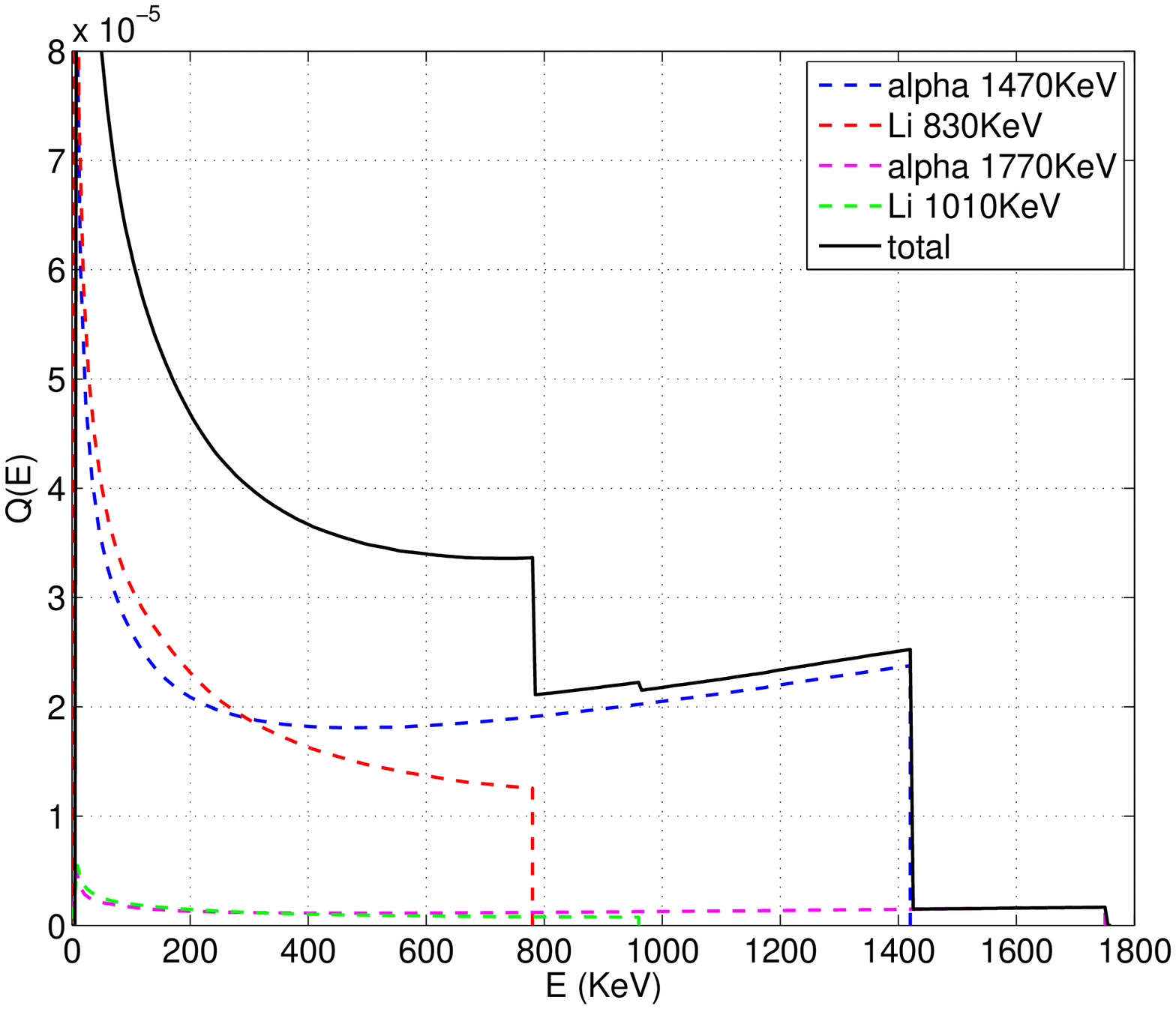}
\includegraphics[width=7.5cm,angle=0,keepaspectratio]{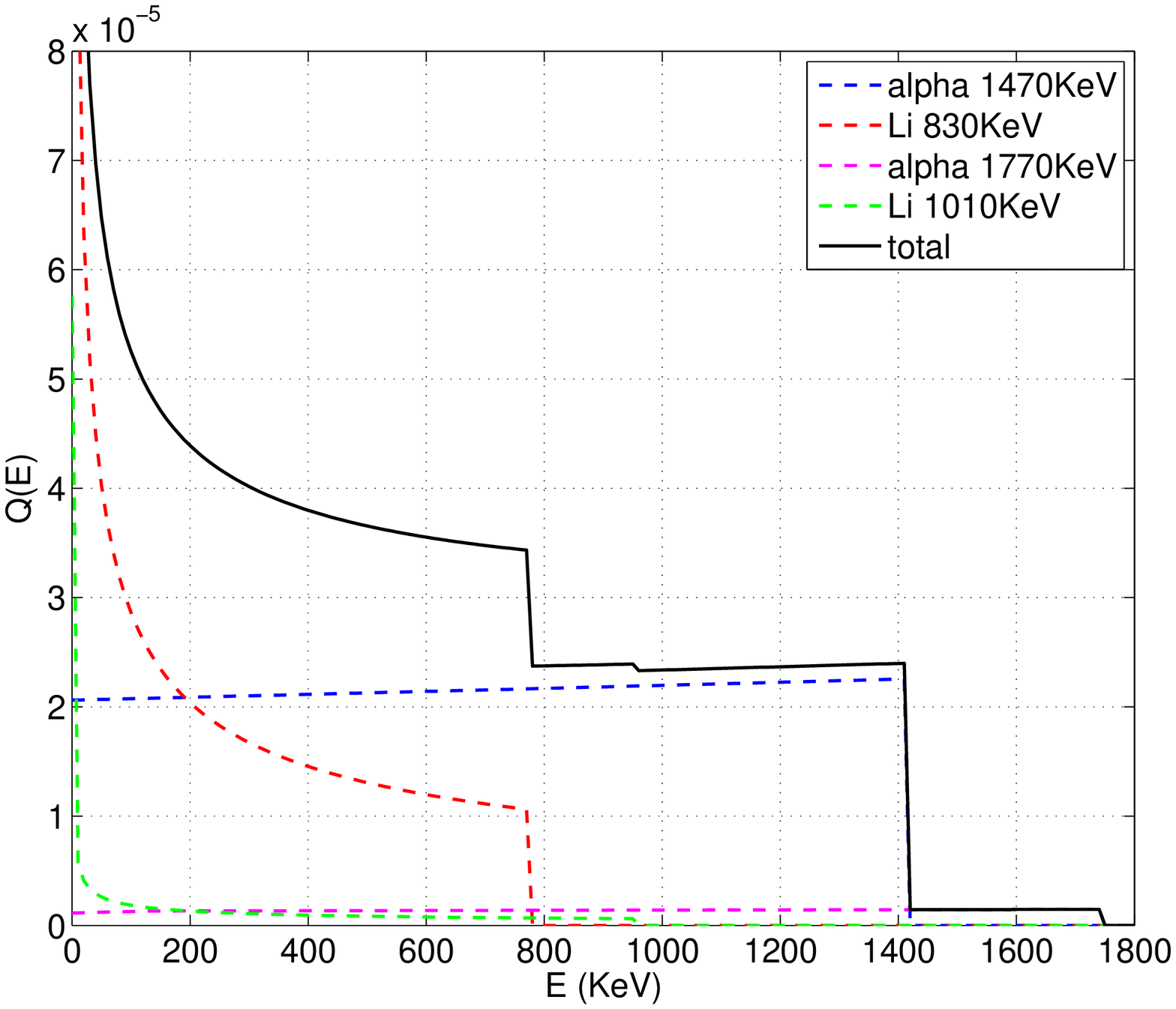}
 \caption{\footnotesize Calculated PHS using SRIM (left) and
 approximated (right) stopping power functions for a single back-scattering layer of $4\, \mu m$
 for $1.8$\AA \, and $90^{\circ}$ incidence.} \label{figPHSST403}
\end{figure}

\begin{figure}[!ht]
\centering
\includegraphics[width=6cm,angle=0,keepaspectratio]{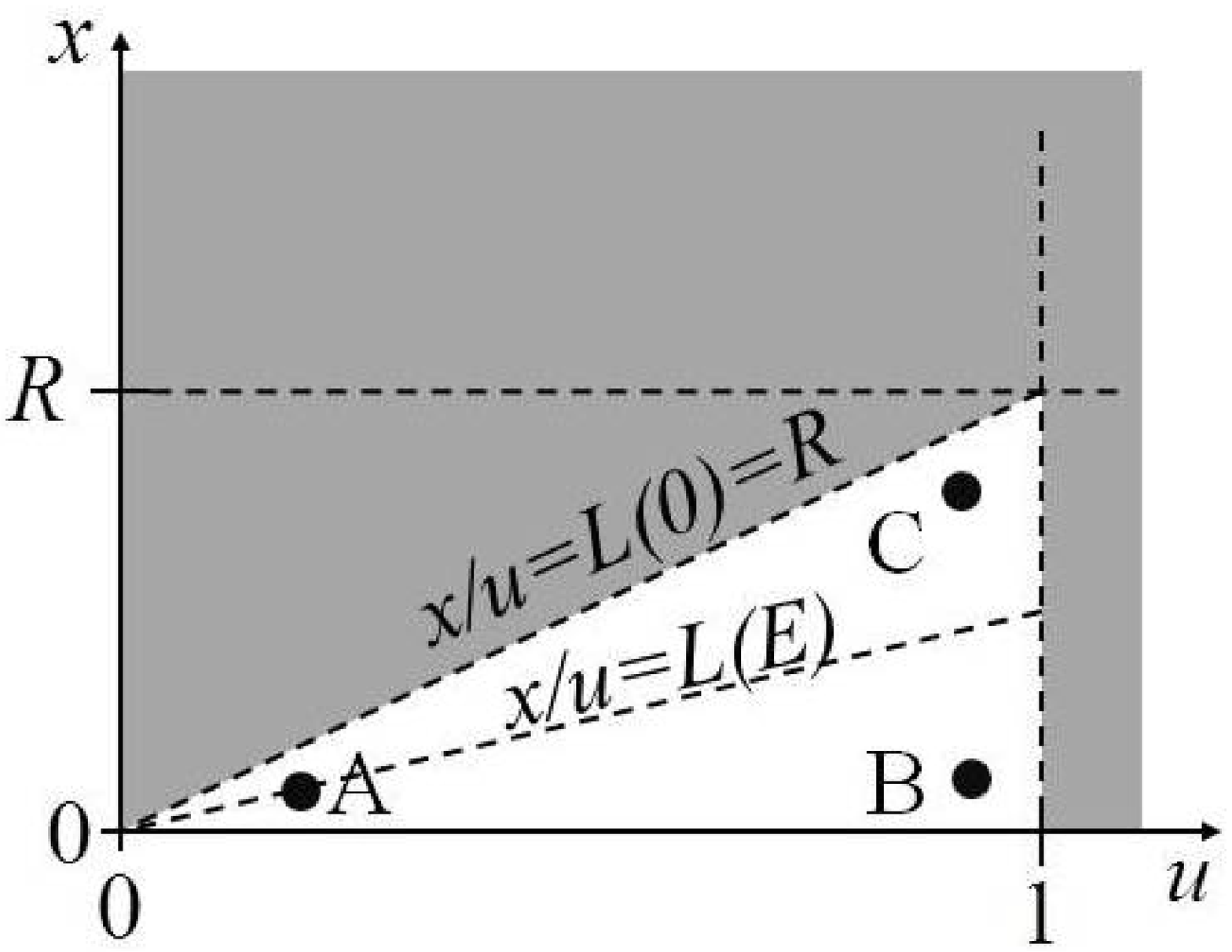}
\includegraphics[width=6cm,angle=0,keepaspectratio]{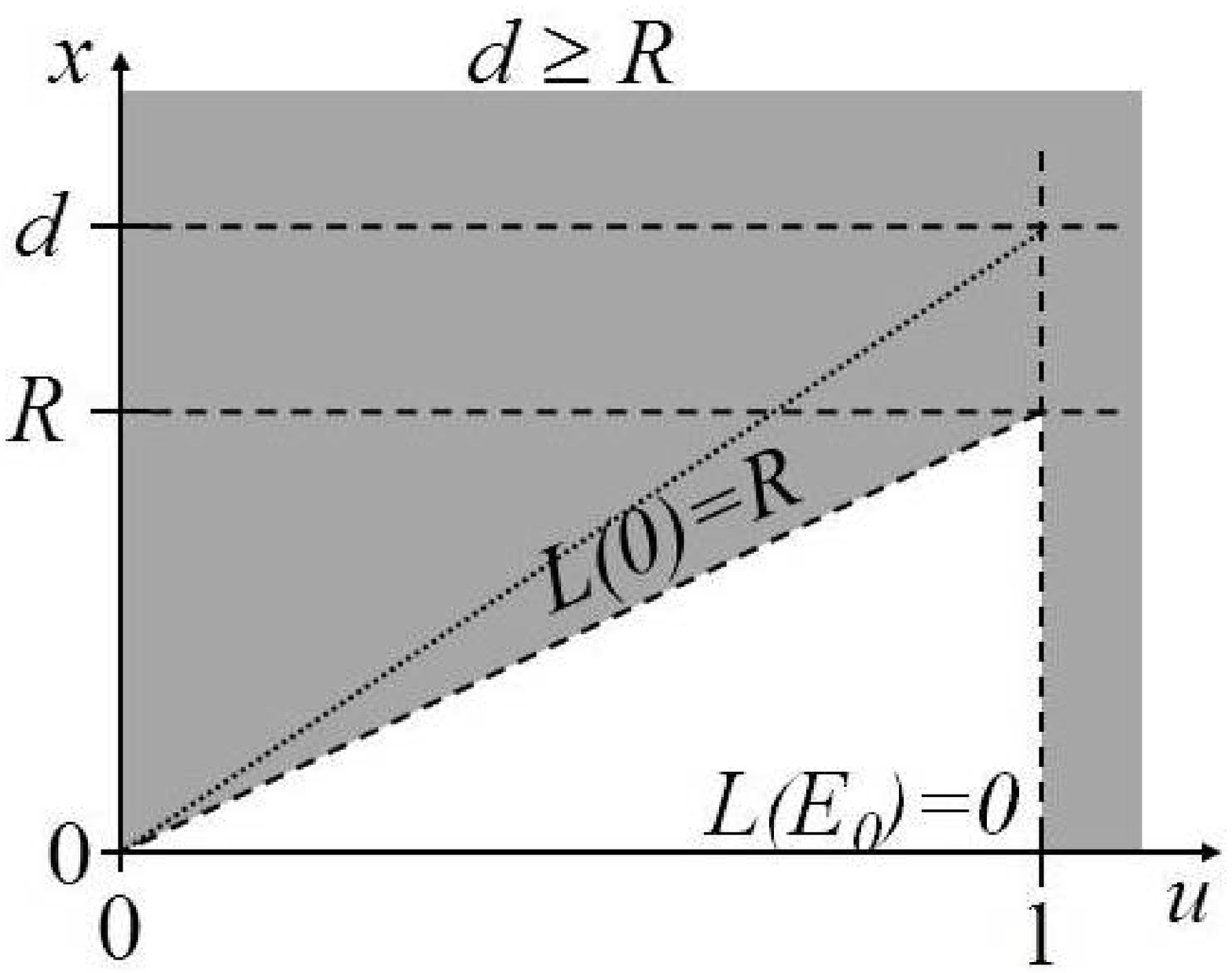}
 \caption{\footnotesize PHS \emph{variable space} sketch (left) and PHS \emph{variable space}
 in the case $d \geq R$ (right).} \label{vs34561}
\end{figure}

A fully analytical result that does not appeal to experimental or
SRIM-calculated stopping power functions, can be useful to
understand the PHS structure and to determine its properties.
\\ The stopping power functions $\frac{dE}{dL}$ can be
approximated by a constant in the case of an $\alpha$-particle and
with a linear dependency in $L$ for a $^7Li$-ion. As a result the
energy dependency as a function of the traveled distance $L$ is
given by:
\begin{equation}\label{eqae9}
E_{\alpha}(L)=\begin{cases} -\frac{E_0}{R} \left( L-R\right) &\mbox{if  \,} L\leq R\\
0 &\mbox{if \, } L > R
\end{cases}
\end{equation}
And equivalently for the $^7Li$-fragment:
\begin{equation}\label{eqae10}
E_{Li}(L)=\begin{cases} \frac{E_0}{R^2}\left( L-R\right)^2 &\mbox{if  \,} L\leq R\\
0 &\mbox{if \, } L > R
\end{cases}
\end{equation}
\\ Where $R$ is the particle range and $E_0$ its initial energy.
\\ In Figure \ref{stpowapprox45} are shown the stopping power
functions $\frac{dE}{dL}$ for $^{10}B$-reaction fragments and their
integral $E(L)$, in the case of using SRIM (solid lines) and in the
case we use the expression displayed in the Equations \ref{eqae9}
and \ref{eqae10} (dashed lines).
By substituting Expressions \ref{eqae9} and \ref{eqae10} into the
Equation \ref{eqae6} we obtain a fully analytical formula for the
PHS. It has to be pointed out that each relation, valid for $L \leq
R$, is valid in the range $E \leq E_0$. Hence Equations \ref{eqae11}
and \ref{eqae12} hold for $E \leq E_0$. The two formulations in
Equation \ref{eqae3} for $L\leq d$ and $L > d$, translate in two
different analytical expressions for $Q(E)$ for $E<E^{*}$ and for
$E\geq E^{*}$, with $d=L(E^{*})$.
\\For the $\alpha$-particle:
\begin{equation}\label{eqae11}
Q(E)\,dE=\begin{cases} \frac{1}{2E_0R\left(1-\frac{E}{E_0}\right)^2}
\left( \frac{1}{\Sigma}-(\frac{1}{\Sigma}+d)e^{-\Sigma
d}\right)\,dE &\mbox{if  \,} E < E_0\left(1-\frac d R\right)\\
\frac{1}{2E_0R\left(1-\frac{E}{E_0}\right)^2} \left(
\frac{1}{\Sigma}-\left(\frac{1}{\Sigma}+R\left(1-\frac{E}{E_0}\right)\right)\cdot
e^{-\Sigma R\left(1-\frac{E}{E_0}\right)}\right)\,dE & \mbox{if \, }
E \geq E_0\left(1-\frac d R\right)
\end{cases}
\end{equation}
\\ Where the relation $E^{*} = E_0\left(1-\frac d R\right)$ is derived from $d=L(E^{*})$.
\\ For the $^7Li$:
\begin{equation}\label{eqae12}
Q(E)\,dE=\begin{cases}
\frac{1}{4E_0R\sqrt{\frac{E}{E_0}}\left(1-\sqrt{\frac{E}{E_0}}\right)^2}
\left( \frac{1}{\Sigma}-(\frac{1}{\Sigma}+d)e^{-\Sigma \cdot
d}\right)\,dE &\mbox{if  \,} E < E_0\left(1-\frac d R\right)^2\\
\frac{1}{4E_0R\sqrt{\frac{E}{E_0}}\left(1-\sqrt{\frac{E}{E_0}}\right)^2}
\left(\frac{1}{\Sigma}-\left(\frac{1}{\Sigma}+\right.\right.\\+\left.\left.R\left(1-\sqrt{\frac{E}{E_0}}\right)\right)
 \cdot e^{-\Sigma
R\left(1-\sqrt{\frac{E}{E_0}}\right)}\right)\,dE &\mbox{if \, } E
\geq E_0\left(1-\frac d R\right)^2
\end{cases}
\end{equation}
\\ Where, again, the relation $E^{*} = E_0\left(1-\frac d R\right)^2$ is
derived from the condition $d=L(E^{*})$.
\\ Figures \ref{figPHSST401}, \ref{figPHSST402} and
\ref{figPHSST403} show the calculated PHS obtained by using the SRIM
stopping power functions and the approximated one displayed in the
Expression \ref{eqae9} and \ref{eqae10} for $0.2\, \mu m$, $1\, \mu
m$ and $4\, \mu m$ respectively, when neutrons hit at $90^{\circ}$
the surface and their wavelength is $1.8$\AA. They show similar
shapes that differ in some points; e.g. focusing on the $1470\,KeV$
$\alpha$-particle, the fact that the approximated $E(L)$ function
(see Figure \ref{stpowapprox45}) differs from the SRIM one at high
$L$ leads to a disappearance of the PHS rise at low energies; it is
clearly visible in Figure \ref{figPHSST403}. \\ We see that as $d$
increases what looked like a single peak splits into two peaks.
While one peak stays constant at the highest fragment energy $E_0$
the second one moves toward lower energies when the layer thickness
increases. This is important when trying to improve the neutron to
gamma-rays discrimination by creating a valley that separates them
in amplitude.
\\ In order to understand the PHS structure, we define the PHS
\emph{variable space}: on the abscissa axis is plotted $u =
\cos(\varphi)$, where $\varphi$ is the angle the fragment has been
emitted with, and, on the ordinates axis, is plotted the neutron
absorption depth $x$. $u \in \left[0,1 \right]$; $x \in \left[0,d
\right]$ if $d<R$ or $x \in \left[0,R \right]$ if $d \geq R$ because
a neutron can only be converted inside the layer and, on the other
hand, if a neutron is converted too deep into the layer, i.e. $x>R$
no fragments can escape whatever the emission angle would be. In
Figure \ref{vs34561}, on its left, the variable space is shown; an
event in the A-position would be a fragment that was generated by a
neutron converted at the surface of the layer and escapes the layer
at grazing angle. An event in the position B represents a fragment
that escapes orthogonally the surface and its neutron was converted
at the surface. An event in C means a neutron converted deep into
the layer with an orthogonal escaping fragment. The straight line
$x=L(E) \cdot u$ characterizes the events with identical escape
energy $E$, that contribute to the same bin in the PHS. The straight
line characterized by $x=R \cdot u$ is the horizon for the particles
that can escape the layer and release some energy in the gas volume.
To be more precise events that give rise to the zero energy part of
the PHS lie exactly on the line $\frac x u =L(E=0)=R$ because they
have traveled exactly a distance $R$ in the converter material. On
the other hand, events that yield almost the full particle energy
$E_0$, will lie on the line identified by $\frac x u=L(E=E_0)=0$.
\\ The events that generate the PHS have access to a
region, on the variable space, identified by a triangle below the
straight line $x=R \cdot u$ (see Figure \ref{vs34561}).
\begin{figure}[!ht]
\centering
\includegraphics[width=6cm,angle=0,keepaspectratio]{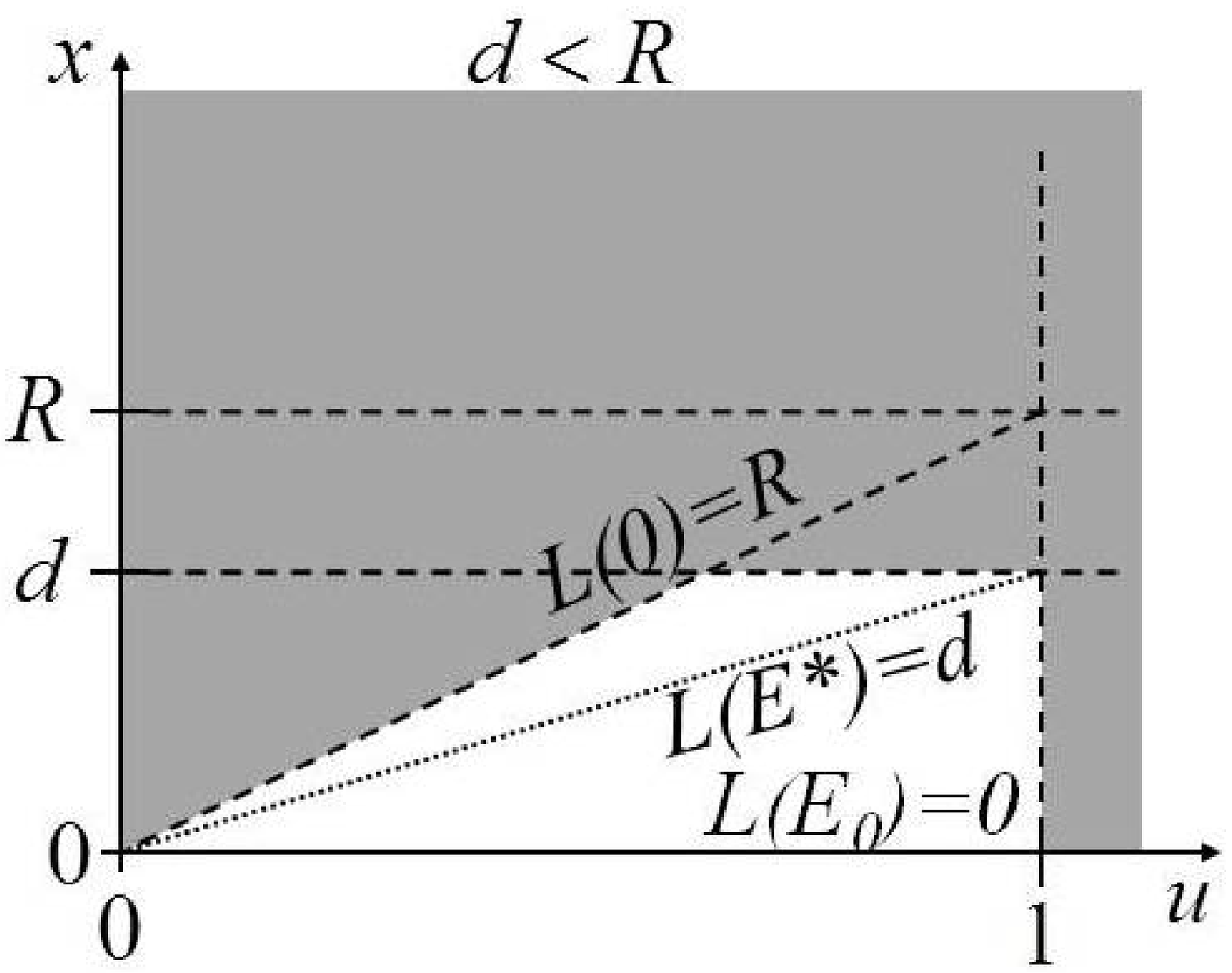}
\includegraphics[width=6cm,angle=0,keepaspectratio]{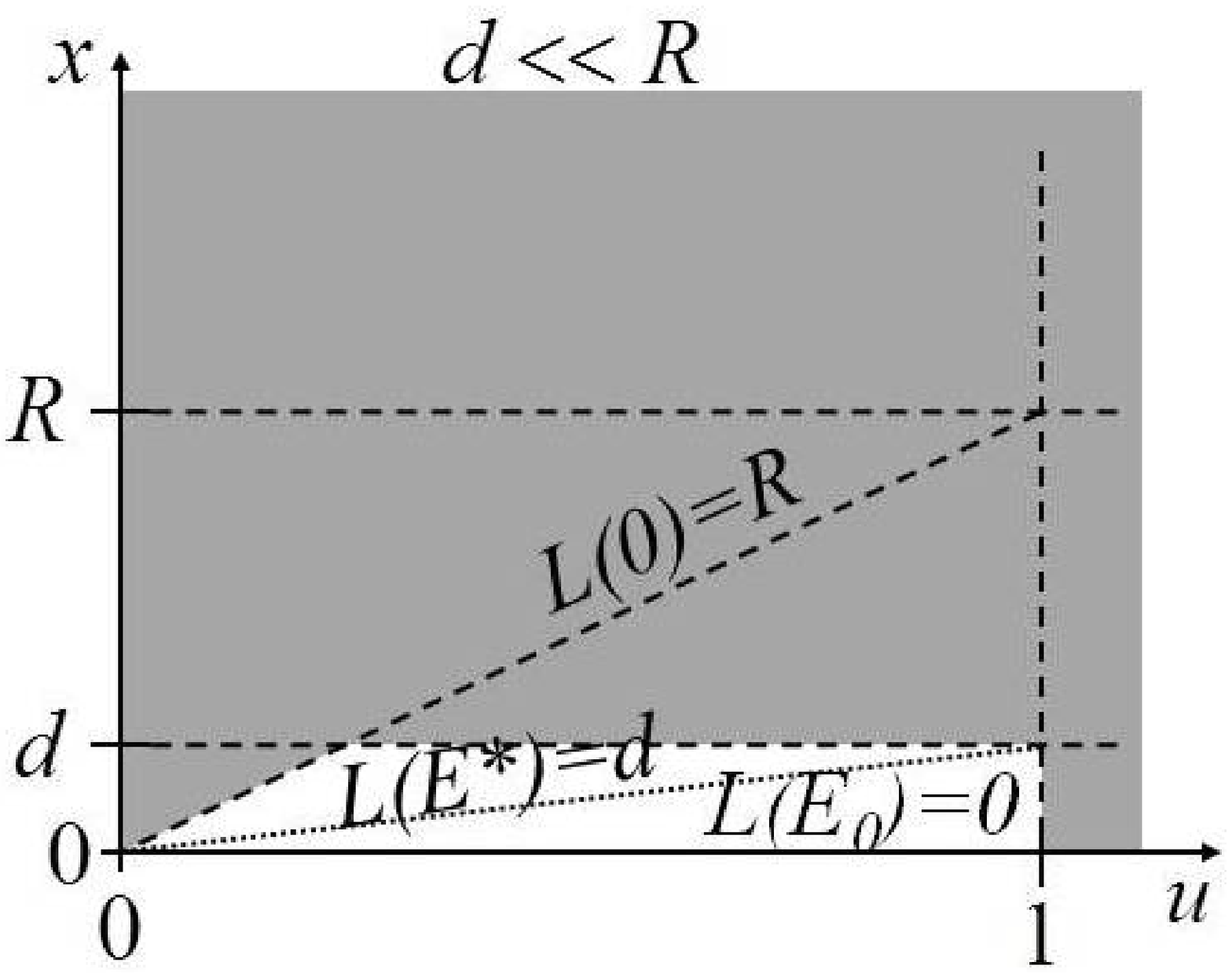}
 \caption{\footnotesize PHS \emph{variable space} in the cases $d < R$ (left)
 and $d << R$ (right).} \label{vs34562}
\end{figure}
\\ If $d > R$ the variable $x$ can explore the interval
$x \in \left[0,R \right]$. This is the case of the PHS in Figure
\ref{figPHSST403}, where $d=4\,\mu m$, $R_{Li(830KeV)}=1.7\, \mu m$
and $R_{\alpha(1470KeV)}=3.4\, \mu m$. We take the two particles of
the $94\%$ branching ratio of $^{10}B$ reaction as example.
\\ If $d<R$, the variable $x$ can explore the interval $x \in
\left[0,d \right]$ (see Figure \ref{vs34562}), thus the domain is
now a trapezoid. The events near the line $\frac x u =L(E^{*})=d$,
which is the switching condition found in the Equations \ref{eqae11}
and \ref{eqae12}, give rise to a peak because this line has the
\emph{maximum length available}. Thus, we expect a peak in the PHS
around $E^{*}$. This is shown in Figure \ref{figPHSST402} where
$d=1\,\mu m$ and, again, $R_{Li(830KeV)}=1.7\, \mu m$ and
$R_{\alpha(1470KeV)}=3.4\, \mu m$, the peaks that originate from the
condition $\frac x u =L(E^{*})=d$ for the two particles are
indicated by the arrows.
\\ If $d\geq R$, the peak occurs for $L(E^{*})=R$, that is, zero
energy. This is problematic for $\gamma$-ray to neutron
discrimination.
\\ If $d<<R$, the variable space is compressed and the
straight lines identified by $\frac x u =L(E^{*})=d$ and $\frac x u
=L(E)=0$ become more and more similar. The two peaks approach and,
the more $d$ is negligible compared to $R$, the more the two peaks
appear as one single peak (see Figure \ref{figPHSST401}).
\\ If we want to avoid a strong presence of neutrons in the low
energy range of the PHS, where we know the $\gamma$-rays
contamination is strong, it is important to try to get the second
peak higher than the energy threshold ($E_{Th}$). This implies that
the thickness $d$ of any layer in the detector should obey
$d<L(E_{Th})$ for the $L$ corresponding to the particle with the
smallest range. This can be a contradictory requirement with
efficiency optimization in which case a compromise between
$\gamma$-rays rejection and efficiency has to be found.

\subsection{Transmission mode}
Equations for transmission mode can be calculated in the same way
they have been determined for back-scattering mode by substituting
$x$ with $d-y$ in the Expression \ref{eqae1} and $x$ with $y$ in the
expression $\delta\left(\frac x u -L\right)$. As a result we obtain:
\begin{equation}\label{eqae13}
P(L)\,dL = \begin{cases} \frac{1}{2 L^2}
\left(\frac{1}{\Sigma}e^{-\Sigma \cdot
d}+(L-\frac{1}{\Sigma})e^{-\Sigma \cdot (d-L)}\right)\,dL &\mbox{if
\,} L \leq d\\ \frac{1}{2 L^2} \left(\frac{1}{\Sigma}e^{-\Sigma
\cdot d}+(d-\frac{1}{\Sigma})\right)\,dL &\mbox{if \, } L > d
\end{cases}
\end{equation}
Hence, $Q(E)dE$ can be calculated as shown already in Section
\ref{backscatt678}. \\ However the same conclusions can be drawn
concerning the qualitative aspects of the PHS, especially the
position of the two peaks.

\section{Conclusions}
We demonstrated that the sputtering technique, suited to make blades
with equal coating thickness on both sides of the substrate, is
well-adapted to make efficiency optimized blades when the substrate
effect can be neglected for any given neutron wavelength and for any
incidence angle distribution. Moreover, this result is also valid
for a multi-layer detector where several blades are arranged in
cascade.
\\ Analytical formulae have been derived in order to optimize the
coating thicknesses of blades in single-blade and in multi-layer
detectors.
\\ The blade-by-blade optimization in the case of a multi-layer detector for a
single neutron wavelength can achieve a few percent more efficiency
over the same blade optimization but this can lead to several blades
less in the detector. Moreover, in the case of a distribution of
wavelengths, the suited optimization from a distribution does not
give important improvements in the overall efficiency compared with
a monochromatic optimization done for the barycenter of the
distribution. On the other hand, the optimization of the efficiency
for a neutron wavelength distribution is often more balanced between
short and long wavelengths than the barycenter optimization.
\\ We have demonstrated that for our model the analytical
expression for the PHS is in a good agreement with measurements.
Moreover, thanks to this model, we understood the overall shape of
the PHS which can be important if one wants to improve the
$\gamma$-ray to neutron discrimination in neutron detectors.

\section{Outlook}
Even though the substrate effect can be neglected in most cases when
dealing with a small amount of blades, its effect in a multi-layer
detector can strongly differ from the results obtained for the ideal
case of completely transparent substrate. A further step is to take
its effect into account.

\appendix
\section{Formulae in \cite{gregor}}\label{app1}
The relations between the formulae in \cite{gregor} and the
expression used in this paper are the following:
\begin{itemize}
    \item the particles ranges $L$ are denoted by $R$;
    \item the branching ratios of the $^{10}B$
reaction (expressed by $F_p$) are $F_1=0.94$ and $F_2=0.06$;
\item the thickness of the layer is $d=D_F$.
\end{itemize}
\noindent Hence, the relation between the expressions \ref{eqac1},
\ref{eqac2}, \ref{eqac4} and the formulae in \cite{gregor} is:
\begin{equation}\label{expla78}
\begin{aligned}
\varepsilon_T(d_T)&=0.94\cdot\varepsilon_{T}(R^{94\%}_1,R^{94\%}_2)+0.06\cdot\varepsilon_{T}(R^{6\%}_1,R^{6\%}_2)=\\
&=S_1(D_F,L_1,0.94)+S_1(D_F,L_2,0.94)+S_2(D_F,L_1,0.06)+S_2(D_F,L_2,0.06)
\end{aligned}
\end{equation}
\\ Valid for both equations ($18a$) ($D_F \leq L_i$) and ($18b$) ($D_F
> L_i$) in Section $4.2$ of \cite{gregor}. In the case of the back-scattering
mode, equations ($25a$) and ($25b$) in \cite{gregor}, we consider
one layer of converter and we replace $\varepsilon_{T}(d_T)$ into
$\varepsilon_{BS}(d_{BS})$ in the expression for back-scattering.
\\ \noindent In a different way, for both equations ($18a$) and ($18b$), we can also write:
\begin{equation}
F_p \cdot \varepsilon_{T}=S_p(D_F,L_1,F_p)+S_p(D_F,L_2,F_p)
\end{equation}

\section{Formulae in \cite{salvat}}\label{app2}
The relations between the formulae in \cite{salvat} and the
expression used in this paper are the following:
\begin{itemize}
    \item the macroscopic cross-section ($\Sigma$) is expressed in terms of mean free path $l=\frac 1
    \Sigma$;
    \item the variable $u$ is denoted by its cosine $u=\cos(\theta)$;
\end{itemize}
The formulae $(4)$ in \cite{salvat} corresponds to the Equation
\ref{eqae1}, unless a factor $\frac 1 2$, where $l=\frac 1 \Sigma$.
\\ The formulae $(3)$ in \cite{salvat} corresponds to the Equation
\ref{eqae3}.

\acknowledgments The authors would like to thank J. Correa and A.
Khaplanov for the data and the Thin Film Physics Division -
Link\"{o}ping University, (Sweden) - especially C. H\"{o}glund, -
for the coatings and B. Gu\'erard, thesis advisor of one of the
authors (F. P.), to have given the opportunity to have worked on
this subject.


\begin{thebibliography}{9}

\bibitem{jonisorma}
J. Birch et al., \emph{$^{10}B_4C$ Multi-Grid as an Alternative to
$^{3}He$ for large area neutron detectors}, IEEE T. Nucl. Sci.,
Volume PP, Issue 99, 17 January 2013, Pages 1-8, ISSN 0018-9499,
\href{http://dx.doi.org/10.1109/TNS.2012.2227798}
{10.1109/TNS.2012.2227798}.

\bibitem{kleinjalousie} M. Henske et al., \emph{The 10B based Jalousie neutron detector $-$ An alternative
for 3He filled position sensitive counter tubes}, Nucl. Instrum.
Meth. A, Volume 686, 11 September 2012, Pages 151-155, ISSN
0168-9002, \href{http://dx.doi.org/10.1016/j.nima.2012.05.075}
{10.1016/j.nima.2012.05.075}.

\bibitem{buff3} J.C. Buffet et al., \emph{Study of a 10B-based Multi-Blade
detector for Neutron Scattering Science}, IEEE T. Nucl. Sci.
Conference Record - Anaheim, 2012.

\bibitem{lacy1}
J. L. Lacy et al., \emph{Boron-coated straws as a replacement for
3He-based neutron detectors}, Nucl. Instrum. Meth. A, Symposium on
Radiation Measurements and Applications (SORMA) XII 2010, Volume
652, Issue 1, 2011, Pages 359-363, ISSN 0168-9002,
\href{http://dx.doi.org/10.1016/j.nima.2010.09.011}{10.1016/j.nima.2010.09.011}.

\bibitem{kouzes2} R. T. Kouzes et al., \emph{Neutron
detection alternatives to 3He for national security applications},
Nucl. Instrum. Meth. A, Volume 623, Issue 3, 2010, Pages 1035-1045,
ISSN 0168-9002, \href{http://dx.doi.org/10.1016/j.nima.2010.08.021}
{10.1016/j.nima.2010.08.021}.

\bibitem{gebauer1} B. Gebauer et al., \emph{Towards detectors for next generation spallation neutron sources},
Proceedings of the 10th International Vienna Conference on
Instrumentation, Nucl. Instrum. Meth. A, Volume 535, Issues 1-2,
2004, Pages 65-78, ISSN 0168-9002,
\href{http://dx.doi.org/10.1016/j.nima.2004.07.266}
{10.1016/j.nima.2004.07.266}.

\bibitem{athanasiades1} A. Athanasiades et al., \emph{Straw detector for high rate, high resolution neutron imaging},
Nuclear Science Symposium Conference Record, 2005 IEEE, Volume 2,
Pages 623-627, 10.1109/NSSMIC.2005.1596338.

\bibitem{gregor2} D.S. McGregor et al., \emph{Semi-insulating bulk GaAs as a
semiconductor thermal-neutron imaging device}, Nucl. Instrum. Meth.
A, Volume 380, Issues 1-2, 1996, Pages 271-275, ISSN 0168-9002,
\href{http://dx.doi.org/10.1016/S0168-9002(96)00347-6}
{10.1016/S0168-9002(96)00347-6}.

\bibitem{tsorbatzoglou1} K. Tsorbatzoglou et al., \emph{Novel and efficient
10B lined tubelet detector as a replacement for 3He neutron
proportional counters}, Symposium on Radiation Measurements and
Applications (SORMA) XII 2010, Nucl. Instrum. Meth. A, Volume 652,
Issue 1, 2010, Pages 381-383, ISSN 0168-9002,
\href{http://dx.doi.org/10.1016/j.nima.2010.08.102}
{10.1016/j.nima.2010.08.102}.

\bibitem{lacy2} J. L. Lacy et al., \emph{One meter square high rate neutron imaging panel based on
boron straws}, Nuclear Science Symposium Conference Record
(NSS/MIC), 2009 IEEE, Pages 1117-1121, ISSN 1095-7863,
\href{http://ieeexplore.ieee.org/stamp/stamp.jsp?tp=&arnumber=5402421&isnumber=5401554}
{10.1109/NSSMIC.2009.5402421}.

\bibitem{gregor}
D.S. McGregor et al., \emph{Design considerations for thin film
coated semiconductor thermal neutron detectors$-I$: basics regarding
alpha particle emitting neutron reactive films}, Nucl. Instrum.
Meth. A, Volume 500, Issues 1-3, 11 March 2003, Pages 272-308, ISSN
0168-9002, \href{http://dx.doi.org/10.1016/S0168-9002(02)02078-8}
{10.1016/S0168-9002(02)02078-8}.

\bibitem{salvat}
D.J. Salvat et al., \emph{A boron-coated ionization chamber for
ultra-cold neutron detection}, Nucl. Instrum. Meth. A, Volume 691, 1
November 2012, Pages 109-112, ISSN 0168-9002,
\href{http://dx.doi.org/10.1016/j.nima.2012.06.041}
{10.1016/j.nima.2012.06.041}.

\bibitem{carina}
C. H\"{o}glund et al., \emph{$B_4C$ thin films for neutron
detection}, J. Appl. Phys., Volume 111, Issue 10, 23 May 2012, Pages
10490-8, ISSN 0168-9002,
\href{http://link.aip.org/link/?JAP/111/104908/1}{10.1063/1.4718573}.

\bibitem{sears} V. F. Sears, \emph{Neutron scattering lengths and cross sections - Special
Feature}, Neutron News, Volume 3, Issue 3, 1992, Pages 29-37.

\bibitem{buff2}
T. Bigault et al., \emph{10B multi-grid proportional gas counters
for large area thermal neutron detectors}, Neutron News, Volume 23,
Issue 4, 2012, Pages 20-25,
\href{http://www.tandfonline.com/doi/abs/10.1080/10448632.2012.725329}{10.1080/10448632.2012.725329}.

\bibitem{wang1} Z. Wang et al., \emph{Multi-layer boron thin-film
detectors for neutrons}, NNucl. Instrum. Meth. A, Volume 652, Issue
1, 1 October 2011, Pages 323-325, ISSN 0168-9002,
\href{http://dx.doi.org/10.1016/j.nima.2011.01.138}{10.1016/j.nima.2011.01.138}.

\bibitem{kleincascade} M.Klein et al., \emph{CASCADE, neutron detectors for highest
count rates in combination with ASIC/FPGA based readout
electronics}, Nucl. Instrum. Meth. A, VCI 2010 Proceedings of the
12th International Vienna Conference on Instrumentation, Volume 628,
Issues 1, 2011, Pages 9-18, ISSN 0168-9002,
\href{http://dx.doi.org/10.1016/j.nima.2010.06.278}{10.1016/j.nima.2010.06.278}.

\bibitem{sri}
J.F. Ziegler et al.,  \emph{SRIM - The stopping and range of ions in
matter (2010)}, Nucl. Instrum. Meth. B, Volume 268, 2010, Pages
1818-1823, \href{http://dx.doi.org/10.1016/j.nimb.2010.02.091}
{10.1016/j.nimb.2010.02.091}.

\end{thebibliography}
\end{document}